\documentclass[12pt]{iopart}
\pdfoutput = 1
\usepackage{graphicx}
\usepackage{iopams}
\usepackage{float}

\usepackage{epsfig}

\begin{document}

\title{Hyperuniformity in point patterns and two-phase random heterogeneous media}


\author{Chase E Zachary$^1$ and Salvatore Torquato$^2$}
\address{$^1$ Department of Chemistry, Princeton University, Princeton, New Jersey 08544, USA}
\ead{czachary@princeton.edu}
\address{$^2$ Department of Chemistry, Princeton University, Princeton, New Jersey 08544, USA;\\
Program in Applied and Computational Mathematics, Princeton University, Princeton, New Jersey 08544, USA;\\
Princeton Institute for the Science and Technology of Materials, Princeton University, Princeton, New Jersey 08544, USA;\\
Princeton Center for Theoretical Science, Princeton University, Princeton, New Jersey 08544, USA; and
School of Natural Sciences, Institute for Advanced Study, Princeton, New Jersey 08540, USA}
\ead{torquato@princeton.edu}

\begin{abstract}
Hyperuniform point patterns are characterized by vanishing infinite-wavelength density fluctuations
and encompass all crystal structures, certain quasi-periodic systems, and special disordered point patterns [S.~Torquato 
and F.~H.~Stillinger, Phys.~Rev.~E \textbf{68}, 041113 (2003)].
This article generalizes the notion of hyperuniformity to include also two-phase random heterogeneous media.  Hyperuniform random media 
do not possess infinite-wavelength volume fraction fluctuations, implying that the variance in the 
local volume fraction in an observation window decays faster than the reciprocal window volume as the window size increases.  
For microstructures of impenetrable and penetrable spheres, we derive an upper bound on the asymptotic coefficient
governing local volume fraction fluctuations in terms of the corresponding quantity describing the variance in the local number density (i.e., number variance).  
Extensive calculations of the asymptotic number variance coefficients are performed for a number 
of disordered (e.g., quasiperiodic tilings, classical ``stealth'' disordered
ground states, and certain determinantal point processes), quasicrystal, and ordered (e.g., Bravais
and non-Bravais periodic systems) hyperuniform structures in various Euclidean space dimensions, and 
our results are consistent with a quantitative order metric characterizing the degree of hyperuniformity.  We also present corresponding 
estimates for the asymptotic local volume fraction coefficients for several lattice families.
Our results have interesting implications for a certain problem in number theory.
\end{abstract}


\maketitle

\section{Introduction}

\begin{figure}[!htp]
\centering
\includegraphics[width=0.45\textwidth]{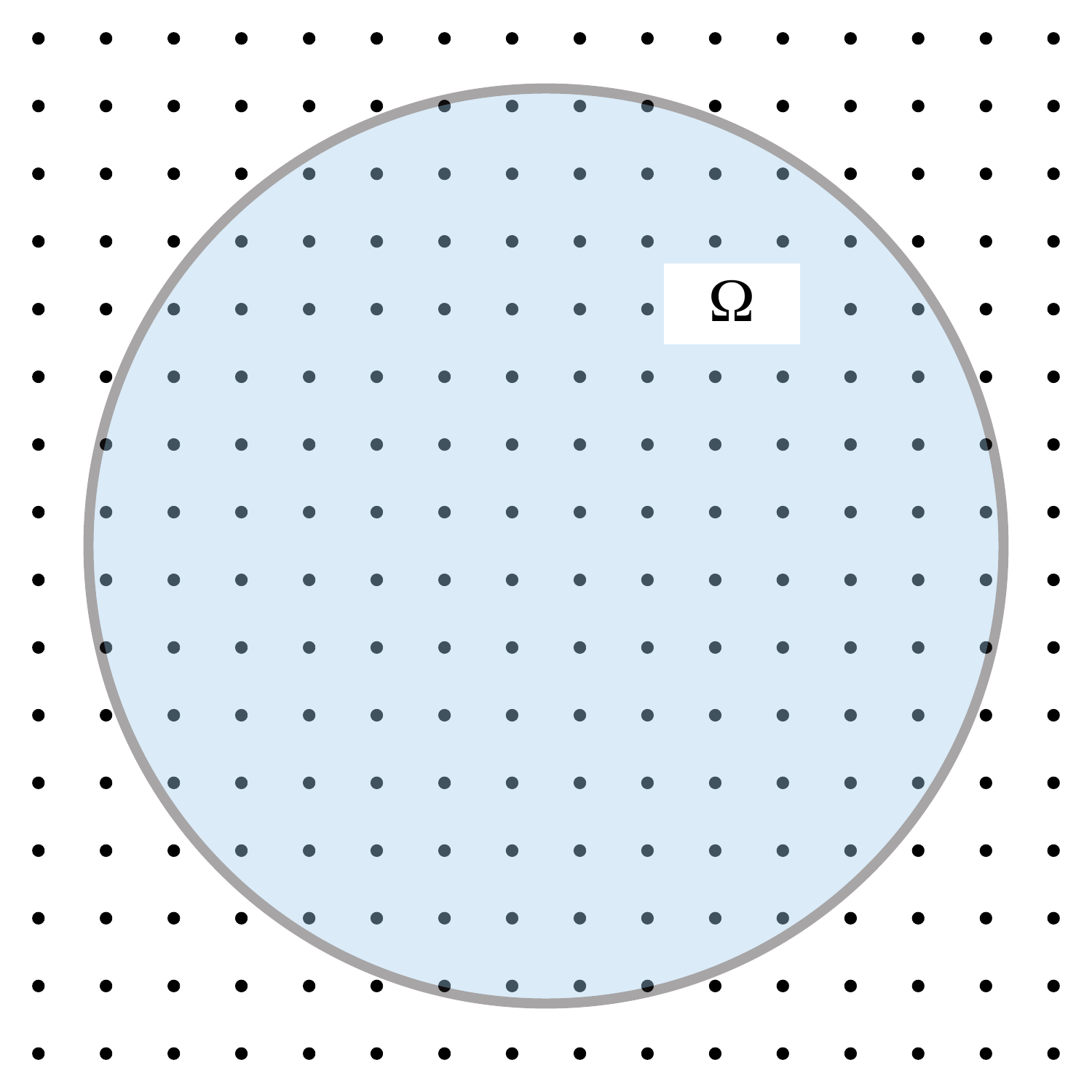}
\includegraphics[width=0.45\textwidth]{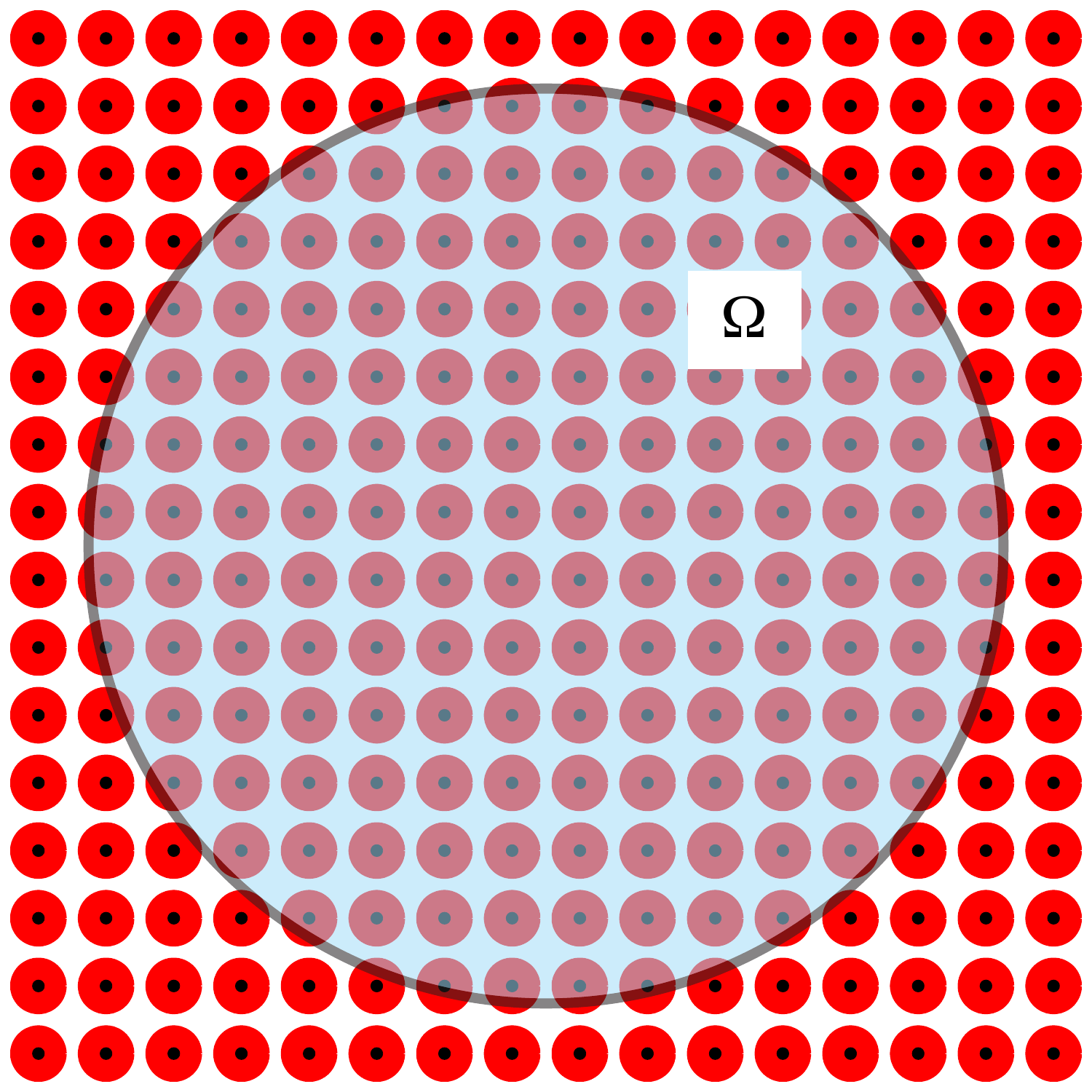}
\includegraphics[width=0.45\textwidth]{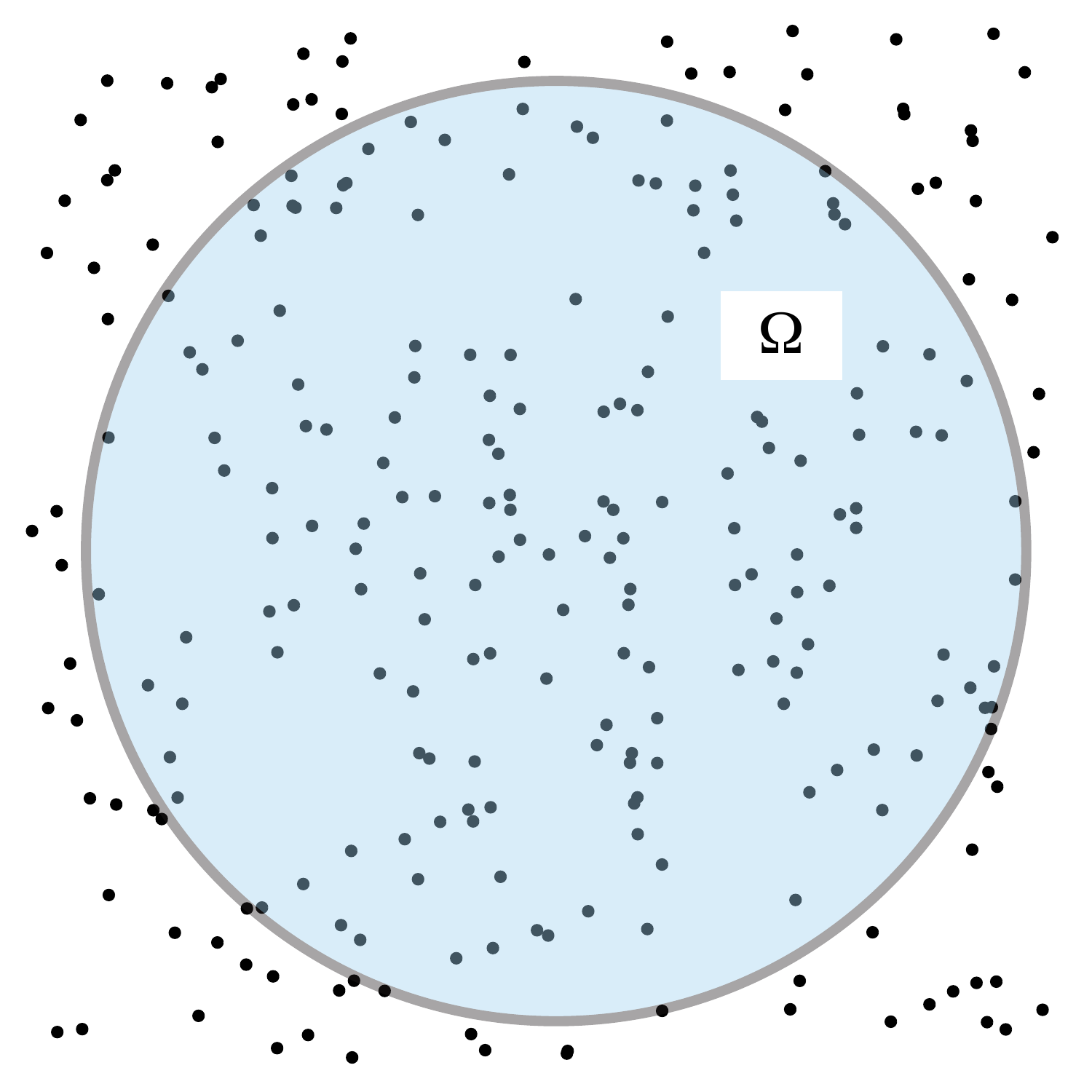}
\includegraphics[width=0.45\textwidth]{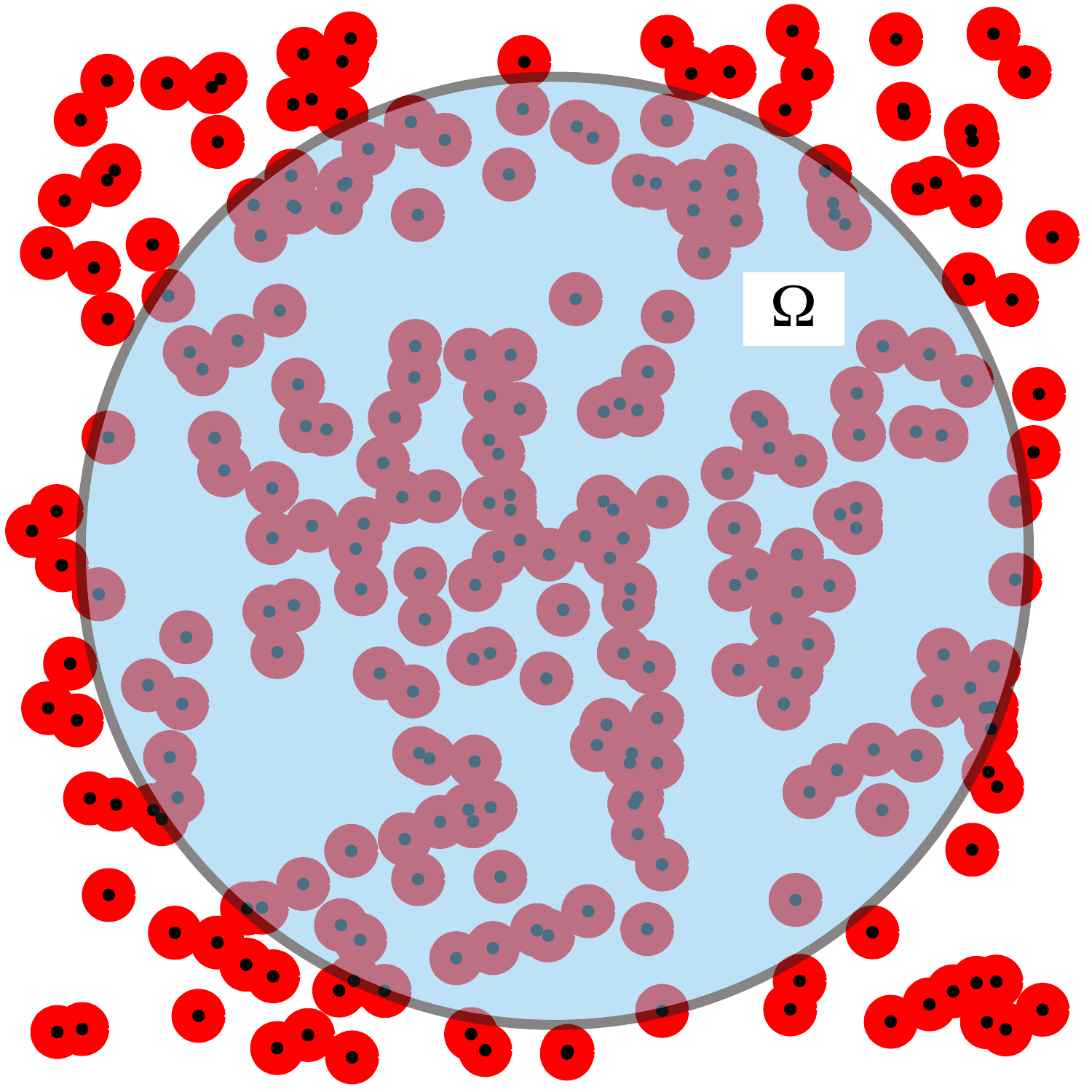}
\caption{\emph{Upper:}  Schematics indicating an observation window $\Omega$ for a periodic point pattern (\emph{left}) and a periodic heterogeneous medium (\emph{right})
obtained by decorating the point pattern with circles.
\emph{Lower:}  Schematics for a disordered point pattern (\emph{left}) and the corresponding disordered heterogeneous medium (\emph{right}).}\label{cartoon}
\end{figure}

Characterizing the local fluctuations in a many-body system represents a fundamental problem in the physical and biological sciences \cite{HaMc86}.
Examples include the large-scale structure of the Universe \cite{PiGaLa02}, 
the structure and collective motion of grains in vibrated 
granular media \cite{WaHa96}, and the structure of living cells \cite{WaYaBaBa02}.  In each of these
cases, one is interested in characterizing the variance in the local number of points of a general point pattern (henceforth known as 
the number variance), and this problem extends naturally to 
higher dimensions with applications to number theory \cite{Ke48, KeRa53} and integrable quantum systems \cite{BlDyLe93,ToScZa08, ScZaTo09}.  Of particular 
importance in this regard is the notion of \emph{hyperuniformity} in a point pattern, in which the number variance in some local observation window 
grows more slowly than the mean as the size of the window increases \cite{ToSt03}; 
such systems are also known as \emph{superhomogeneous} \cite{PiGaLa02}. Figure \ref{cartoon} shows typical periodic and nonperiodic point patterns 
along with the corresponding observation windows.
Torquato and Stillinger \cite{ToSt03} have provided a rigorous characterization of both periodic and nonperiodic hyperuniform point patterns, which do not possess 
infinite-wavelength local density fluctuations.  They have also computed the asymptotic number variance
coefficients for a number of common lattices, including the square, triangular, FCC, and BCC crystal structures.    
Further examples of hyperuniform systems include all periodic point patterns, certain quasi-periodic systems, and 
special disordered point patterns.  We remark that a mathematical (or \emph{Bravais}) \emph{lattice}
is a system of points in Euclidean space with a set of minimal basis vectors $\left\{\mathbf{e}_i\right\}$ such that any point in the lattice can be 
generated by taking integral linear combinations of the basis vectors.  Other periodic structures can be formed by including more than one point in a fundamental cell, and 
we will occasionally refer to such structures as nonlattices.  Unless otherwise noted, our usage of the term \emph{lattice} will exclusively apply to Bravais lattices.

Here we generalize the notion of hyperuniformity to heterogeneous media,  
wherein one characterizes the local fluctuations in the volume fraction.
A \emph{heterogeneous medium}, also known in the mathematics literature as a \emph{random set},
is a partition of $d$-dimensional Euclidean space $\mathbb{R}^d$ into disjoint regions (phases) with interfaces that are known only probabilistically.
Throughout this paper we will consider two-phase heterogeneous media.
Microstructural information for the system is contained in the indicator function $I^{(i)}$ for each phase $i$, defined by:
\begin{eqnarray}
I^{(i)}(\mathbf{r}) = \left\{
\begin{array}{lr}
1, \quad \mathbf{r} \in \mbox{phase } i\\
0, \quad \mbox{else}.
\end{array}\right.
\end{eqnarray}
Of particular
importance is the so-called two-point probability function $S_2(\mathbf{r}_1, \mathbf{r}_2)$ \cite{St95, ToSt82, ToSt83, torquato2002rhm}, which is defined by:
\begin{eqnarray}
S_2(\mathbf{r}_1, \mathbf{r}_2) = \langle I(\mathbf{r}_1) I(\mathbf{r}_2)\rangle,
\end{eqnarray}
where we have implicitly chosen some reference phase in the system.  Note that since one can always construct a heterogeneous medium 
by ``decorating'' an appropriate point process (e.g., circumscription with hexagons or spheres; see Figure \ref{cartoon}), characterizing the 
statistical properties of a heterogeneous medium represents a more general problem than the study of point patterns; namely, the 
latter systems are recovered as a special limit of the former.  
It is therefore intuitive that any physical application of point processes is also 
important for heterogeneous media.  Applications of the hyperuniformity problem for such systems include scattering 
by heterogeneous media \cite{DeAnBr57}, transport through composites and porous media \cite{ToLa91}, 
the study of noise and granularity of photographic images \cite{Ba64, LuTo90b}, identifying properties of 
organic coatings \cite{FiKuBi92}, and the fracture of composite materials \cite{To00}.  In fact, it is generally true that any statistical description 
of the microstructure of a heterogeneous material provides nontrivial information about a wide variety of effective properties of that material in a homogenized sense 
\cite{torquato2002rhm}.

The local volume fraction $\tau$ of a heterogeneous medium is defined as the fraction of the volume of an observation window
covered by a particular phase (see Figure \ref{cartoon}).  For the remainder of this paper we consider spherical observation windows, but our results
are easily extended to windows of arbitrary geometry \cite{ToSt03}.
Lu and Torquato \cite{LuTo90} have studied the variance $\sigma^2_{\tau}$ of $\tau$, which they refer to as the 
\emph{coarseness} after suitable scaling, and have shown in the homogeneous case that it is related to the 
two-point probability function according to:
\begin{eqnarray}
\sigma^2_{\tau} = \frac{1}{v(R)} \int_{\mathbb{R}^d} S_2(\mathbf{r}) \alpha(r; R) d\mathbf{r} - \phi^2,
\end{eqnarray}
where $v(R)$ is the volume of a sphere of radius $R$, $\alpha$ is the so-called scaled intersection volume of two spheres of radius $R$ whose centers are separated by a distance $r$ (defined in more
detail in Section 2), and $\phi$ is the fraction of space in the global medium occupied by the reference phase.  For large
windows, these authors showed that in non-hyperuniform media $\sigma^2_{\tau} \sim [v(R)]^{-1}$. Additionally, Quintanilla and Torquato \cite{QuTo97} 
have determined the cumulative distribution functions
of the local volume fraction for a number of low-dimensional periodic \cite{QuTo99} and random \cite{QuTo97} media; it is known that a 
central-limit result holds for sufficiently large observation windows \cite{QuTo97}.   
We mention here that there is an 
analogous formula for the number variance of a point pattern:
\begin{eqnarray}
\sigma^2_N = \rho v(R)\left[1+\rho\int_{\mathbb{R}^d} h(\mathbf{r}) \alpha(r; R) d\mathbf{r}\right],
\end{eqnarray}
where $h$ is the so-called total correlation function of the point process.  The details behind these relations are provided in Sections 2 and 3,
but it is clear from these remarks that the coarseness of a heterogeneous medium plays a similar role as the local density fluctuations in a point pattern.
Figure \ref{cartoon} compares the hyperuniformity problem for 
point patterns and heterogeneous media in both the periodic and nonperiodic cases.  Unlike the local number density, the local volume
fraction approaches a fixed value $\phi$ on the global scale of the system, meaning that the coarseness must \emph{decay} over large length scales.  The 
analog of hyperuniformity in this instance will involve a decay in the local volume fraction fluctuations that is \emph{faster} than the volume of the observation window;
we quantify this behavior in Section 3 using the asymptotic scaling of the coarseness.

One important implication of this work is the development of a quantitative order metric characterizing the degree of order in a point pattern.  From our experience with low-dimensional systems, it is physically intuitive that Bravais lattices are in some sense ``more ordered'' than nonlattice structures and quasicrystals, which in turn are seemingly more regular than strongly-correlated disordered point patterns.  Furthermore, one also expects that weakly correlated systems are highly disordered.  It is thus an open problem in statistical physics to quantify ``randomness'' in a many-body system with a simple metric that reflects this intuitive ordering.  The observation that hyperuniform point patterns are characterized by a certain regularity in the distribution of points throughout $\mathbb{R}^d$ suggests that quantitative measurements of the local number variance can provide one measure of order in point patterns, and we discuss the construction of such a metric in this work.  

Section 2 reviews the basic notions of stochastic point processes, which are completely characterized by the set of $n$-particle correlation functions, and 
hyperuniformity, which is defined completely by the two-particle information of the point process.  Section 3 presents the corresponding results for two-phase
heterogeneous media, whereby the fluctuations in the local volume fraction are related to two-point microstructural information of the phases.  In general, 
if a heterogeneous medium can be generated from a stochastic point process, even lower-order microstructural information about the system will require a knowledge 
of the full $n$-particle information of the underlying point process.  However, we show in Section 4 that there is a convenient relation between fluctuations 
in the local number density and the corresponding fluctuations in the local volume fraction for microstructures of spherical inclusions, thereby providing
a computationally accessible estimate of the local volume fraction variance.  Section 5 presents explicit calculations of the asymptotic coefficients governing 
the local density fluctuations for a wide range of periodic and disordered point processes in low Euclidean dimensions, and these results translate into the aforementioned quantitative order metric characterizing 
the degree of (dis)order in a given arrangement of points.  We extend these calculations in Section 6 to include the asymptotic properties of these terms for large Euclidean dimensions and also to provide information about the corresponding coefficients for 
the fluctuations in the local volume fraction.  Extensive tables of asymptotic coefficients are provided for periodic and nonperiodic systems in Sections 5 and 6, and the reader who wishes to skip the mathematical 
details of the preceding sections is directed there.
Our results have interesting implications for the minimzers of
the Epstein zeta function, a fundamental problem that arises in number theory.
Concluding remarks are given in Section 7.      

\section{Background and definitions}

\subsection{Point processes and $n$-particle correlation functions}

A \emph{stochastic point process} in $\mathbb{R}^d$ 
is defined as a mapping from a probability space
to configurations of points $\mathbf{r}_1, \mathbf{r}_2, \mathbf{r}_3\ldots$
in $d$-dimensional Euclidean space $\mathbb{R}^d$. More precisely, let $X$ denote
the set of configurations such that each configuration 
$x \in X$ is a subset of $\mathbb{R}^d$ that satisfies two regularity
conditions: (i) there are no multiple points
($\mathbf{r}_i \neq \mathbf{r}_j$ if $ i\neq j$) and (ii)
each bounded subset of $\mathbb{R}^d$ must contain
only a finite number of points of $x$.
We denote by $N(B)$  the number of points within
$x \cap B$, $B \in \mathcal{B}$, where $\mathcal{B}$ is
the ring of bounded Borel sets in $\mathbb{R}^d$. Thus, we always
have $N(B) < +\infty$ for $B \in \mathcal{B}$, but the
possibility $N(\mathbb{R}^d)= +\infty$ is not excluded.
We note that there exists a minimal $\sigma$-algebra $\mathcal{U}$ of subsets
of $X$ that renders all of the functions $N(B)$ measurable.
Let $(\Omega,\mathcal{F}, \mathcal{P})$ be a probability
space. Any measurable map $x: \Omega \rightarrow X$
is called a stochastic point process \cite{St95,ToSt06}.
Henceforth, we will simply call this map a point process.
Note that this random setting is quite general;
it incorporates cases in which the location of the points are deterministically
known, such as a lattice. 

A point process is completely specified by the countably infinite set of generic $n$-particle probability density functions.  
The \emph{generic $n$-particle probability density function},
denoted by $\rho_n(\mathbf{r}_1,\ldots,\mathbf{r}_n)$, is proportional to the probability density of finding $n$ particles in volume
elements around the given positions $(\mathbf{r}_1, \ldots, \mathbf{r}_n)$, 
irrespective of the remaining $N-n$ particles. 
Specifically, if $P_N(\mathbf{r}_1, \ldots, \mathbf{r}_N)$ is the probability density for finding $N$ particles at positions $(\mathbf{r}_1, \ldots, \mathbf{r}_N)$,
the generic $n$-particle probability density is given by:
\begin{eqnarray}
\rho_n(\mathbf{r}_1, \ldots, \mathbf{r}_n) = \frac{N!}{(N-n)!} \int P_N(\mathbf{r}_1, \ldots, \mathbf{r}_n, \mathbf{r}_{n+1}, \ldots, \mathbf{r}_N) 
\prod_{j=n+1}^{N} d\mathbf{r}_j.
\end{eqnarray}
For a statistically homogeneous point process, the single-particle generic density function is:
\begin{eqnarray}\label{thermlimit}
\rho_1(\mathbf{r}_1)= \rho \equiv \lim_{N, V\rightarrow +\infty} N/V,
\end{eqnarray}
where the limit in (\ref{thermlimit}) defines the \emph{thermodynamic limit}.
This function is proportional to 
the probability density of finding a particle at $\mathbf{r}_1$, also known as the \emph{intensity} of the point process.  
In general the normalization for $\rho_n$ is given by $N!/(N-n)!$, or the number of ways of choosing an ordered subset of $n$ points from a population of size $N$.
For a translationally-invariant and completely uncorrelated point process $\rho_n(\mathbf{r}_1,\ldots, \mathbf{r}_n) = \rho^n$ for all $n$.

We also introduce the \emph{$n$-particle correlation functions} $g_n$, which are defined by:
\begin{eqnarray}
g_n(\mathbf{r}_1,\ldots,\mathbf{r}_n) = \frac{\rho_n(\mathbf{r}_1,\ldots,\mathbf{r}_n)}{\rho^n}.
\end{eqnarray}
Since $\rho_n = \rho^n$ for a completely uncorrelated point process, it follows that deviations of $g_n$ from unity provide a measure of the correlations
between points in a point process.
Of particular interest is the pair correlation function, which for a translationally-invariant point process of intensity $\rho$ can be written as:
\begin{eqnarray}\label{g2detpp}
g_2(\mathbf{r})=\frac{\rho_2(\mathbf{r})}{\rho^2} \qquad (\mathbf{r} = \mathbf{r}_2 - \mathbf{r}_1).
\end{eqnarray}
Closely related to the pair correlation function is the \emph{total correlation function}, denoted by $h$; it is derived from $g_2$
via the equation:
\begin{eqnarray}\label{hdef}
h(\mathbf{r}) = g_2(\mathbf{r}) - 1.
\end{eqnarray}
Since $g_2(r) \rightarrow 1$ as $r\rightarrow +\infty$ ($r = \Vert\mathbf{r}\Vert$) for translationally invariant systems without long-range order, 
it follows that $h(r)\rightarrow 0$ in this limit, meaning that $h$ is generally 
an $L^2$ function, and its Fourier transform is well-defined.  

It is common in statistical mechanics when passing to reciprocal space to
consider the associated \emph{structure factor} $S$, which for a translationally invariant system is defined by:
\begin{eqnarray}\label{Sdef}
S(k) = 1+\rho\hat{h}(k),
\end{eqnarray}
where $\hat{h}$ is the Fourier transform of the total correlation function, $\rho$ is the number density, and $k = \Vert\mathbf{k}\Vert$ is the magnitude of the 
reciprocal 
variable to $\mathbf{r}$.  
We utilize the following definition of the
Fourier transform:
\begin{eqnarray}
\hat{f}(\mathbf{k}) = \int_{\mathbb{R}^d} f(\mathbf{x}) \exp\left[-i(\mathbf{k}, \mathbf{r})\right] d\mathbf{r},
\end{eqnarray}
where $(\mathbf{k}, \mathbf{r}) = \sum_{i=1}^d k_i r_i$ is the conventional Euclidean inner product of two real-valued vectors.  
When it is well-defined, the corresponding inverse Fourier transform is given by:
\begin{eqnarray}
f(\mathbf{r}) = \left(\frac{1}{2\pi}\right)^d \int_{\mathbb{R}^d} \hat{f}(\mathbf{k}) \exp\left[i(\mathbf{k}, \mathbf{r})\right] d\mathbf{k}.
\end{eqnarray}
We remark that for radially-symmetric functions [i.e., $f(\mathbf{r}) = f(\Vert\mathbf{r}\Vert) = f(r)$], the Fourier and inverse Fourier transforms may
be written respectively as follows:
\begin{eqnarray}
\hat{f}(k) &= (2\pi)^{d/2}\int_0^{\infty} r^{d-1} f(r) \frac{J_{(d/2)-1}(kr)}{(kr)^{(d/2)-1}} dr\\
f(r) &= \left(\frac{1}{2\pi}\right)^{d/2} \int_0^{\infty} k^{d-1} \hat{f}(k) \frac{J_{(d/2)-1}(kr)}{(kr)^{(d/2)-1}} dk.
\end{eqnarray}

\subsection{Hyperuniform point processes}

A \emph{hyperuniform} point process has the property that the variance in the number of points in an observation window $\Omega$ grows more slowly than the volume 
of that window.  In the case of a spherical observation window, this definition implies that the local number variance $\sigma^2_N(R)$ grows more slowly than $R^d$ 
in $d$ dimensions, where $R$ is the radius of the observation window.  Torquato and Stillinger \cite{ToSt03} (see also \cite{Le83, LaLi80, MaYa80} for related results) have provided 
an exact expression for the local number variance
of a statistically homogeneous point process in a spherical observation window:
\begin{eqnarray}\label{numvar}
\sigma^2_N(R) = \rho v(R) \left[1+\rho \int_{\mathbb{R}^d} h(\mathbf{r}) \alpha(r; R) d\mathbf{r}\right],
\end{eqnarray}
where $R$ is the radius of the observation window, $v(R)$ is the volume of the window, and $\alpha(r; R)$ is the 
so-called \emph{scaled intersection volume}.  The latter quantity is geometrically defined as the volume of space occupied by the intersection 
of two spheres of radius $R$ separated by a distance $r$ normalized by the volume of a sphere $v(R)$.
We remark that the average number of points in an observation window is $\langle N(R)\rangle = \rho v(R)$
for any statistically homogeneous point process.
 
The scaled intersection volume then  has the support
$[0,2R]$, the range $[0,1]$, and the following alternative integral representation \cite{ToSt06}:
\begin{eqnarray}
\alpha(r;R) = c(d) \int_0^{\cos^{-1}[r/(2R)]} \sin^d(\theta) d\theta,
\label{alpha}
\end{eqnarray}
where $c(d)$ is a $d$-dimensional constant given by
\begin{eqnarray}
c(d)= \frac{2 \Gamma(1+d/2)}{\pi^{1/2} \Gamma[(d+1)/2]}.
\label{C}
\end{eqnarray}
Torquato and Stillinger \cite{ToSt06} found the following  series representation
of the scaled intersection volume $\alpha(r;R)$ for $r \le 2R$
and for any $d$:
\begin{eqnarray}
\fl \alpha(r;R)=1- c(d) x+ 
c(d) \sum_{n=2}^{\infty}
(-1)^n \frac{(d-1)(d-3) \cdots (d-2n+3)}
{(2n-1)[2 \cdot 4 \cdot 6 \cdots (2n-2)]} x^{2n-1},
\label{series}
\end{eqnarray}
where $x=r/(2R)$. For even dimensions, relation (\ref{series}) is
an infinite series, but for odd dimensions, the series truncates such
that $\alpha(r;R)$ is a univariate polynomial of degree $d$. 
It is convenient to introduce a dimensionless density $\phi$, defined by:
\begin{eqnarray}
\phi = \rho v(D/2) = \frac{\rho \pi^{d/2} D^d}{2^d\Gamma(1+d/2)},
\end{eqnarray}
where $D$ is a characteristic microscopic length scale of the system (e.g., the mean nearest-neighbor distance between points).  

Using the expansion (\ref{series}),
the result in (\ref{numvar})
admits the following asymptotic scaling \cite{ToSt03}:
\begin{eqnarray}\label{numasymp}
\sigma_N^2(R) = 2^d \phi\left\{A_N\left(\frac{R}{D}\right)^d + B_N\left(\frac{R}{D}\right)^{d-1} + o\left[\left(\frac{R}{D}\right)^{d-1}\right]\right\},
\end{eqnarray}
where $o(x)$ denotes all terms of order less than $x$.   
Explicit forms for the asymptotic coefficients $A_N$ and $B_N$ are given by:
\begin{eqnarray}
A_N &= 1+\rho\int_{\mathbb{R}^d} h(\mathbf{r}) d\mathbf{r} = \lim_{\Vert\mathbf{k}\Vert\rightarrow 0} S(\mathbf{k})\label{An}\\
B_N &= -\frac{\phi \kappa(d)}{D v(D/2)}\int_{\mathbb{R}^d} h(\mathbf{r}) \Vert\mathbf{r}\Vert d\mathbf{r}\label{Bn},
\end{eqnarray} 
where $\kappa(d) = c(d)/2$ with $c(d)$ defined by (\ref{C}); note that we have taken $R\rightarrow +\infty$ in the integrals, thereby rendering $A_N$ and $B_N$ independent of $R$.
These results are valid for all periodic point patterns (including
lattices), quasicrystals that possess Bragg peaks, and disordered systems in which 
the pair correlation function $g_2$ decays to unity exponentially fast \cite{ToSt03, To09}. 
Any such system with $A_N = 0$ satisfies the requirements for hyperuniformity.  The relation in (\ref{An}) then implies that hyperuniform point patterns
do not possess infinite-wavelength fluctuations in the local number density.  

For disordered hyperuniform systems with a total correlation function $h$ that does not decay to zero exponentially fast, other dependencies of the number variance on $R$ may be observed.
For example, it is known that if the total correlation function $h \sim r^{-(d+1)}$ for large $r$, then $B_N \sim a_0\ln R + a_1$ \cite{ToScZa08}.
Such behavior occurs in maximally random jammed sphere packings \cite{DoStTo05} 
and noninteracting spin-polarized fermion ground states \cite{ToScZa08, ScZaTo09}.
More generally, for any reciprocal power law: 
\begin{eqnarray}
h(r) \sim -\frac{1}{r^{d+\zeta}} \qquad (r\rightarrow +\infty), 
\end{eqnarray}
one can observe a number of 
different types of dependencies of the number variance $\sigma^2_N$ on the window radius $R$:
\begin{eqnarray}
\sigma^2_N(R) \sim \left\{
\begin{array}{lr}
R^{d-1}\ln R, \quad [\zeta = 1]\\
R^{d-\zeta}, \quad [\zeta \in (0, 1)].  
\end{array}\right.
\end{eqnarray}
Note that it is not possible to construct a hyperuniform system for which $\zeta \leq 0$ since the number variance would then grow at least as fast as the volume of the observation window. It is also not possible to choose $\zeta > 1$ since the number variance cannot grow any slower than the surface area of the window \cite{Be87, ToSt03}.  However, there are examples of point patterns that do not possess a surface area term in the number variance (i.e., $B_N = 0$); these systems are known as \emph{hyposurficial} point patterns \cite{ToSt03}.        

Cases where the number variance grows faster than the window volume have been reported on the real line \cite{CoDuHu08}.  These non-hyperuniform systems exhibit what we describe as ``super-Poissonian'' behavior\footnote{Super-Poissonian point processes have also been called ``sub-homogeneous'' in the literature \cite{CoDuHu08}.}
and are similar to thermodynamic critical points \cite{ToSt03}, where the structure factor diverges at the origin.  Note that this behavior implies a scale-invariant structure.  More generally, if a non-hyperuniform point pattern possesses a total correlation function:
\begin{eqnarray}
h(r) \sim \frac{1}{r^{d+\xi}} \qquad (r\rightarrow +\infty),
\end{eqnarray}
then the number variance can exhibit the following asymptotic behaviors:
\begin{eqnarray}  
\sigma^2_N(R) \sim \left\{
\begin{array}{lr}
R^d, \quad (0 < \xi < +\infty)\\
R^d \ln R, \quad (\xi = 0)\\
R^{d+\vert\xi\vert}, \quad (-d < \xi < 0).
\end{array}\right.
\end{eqnarray}
Note that the integrability requirement of the function $h(r)\alpha(r; R)$ implies that the number variance cannot grow any faster than the square of the window volume.


We remark that the problem of minimizing the number variance can be expressed in terms of finding the ground state of a 
pair interaction with compact support; namely, by invoking a volume-average interpretation of the number variance problem valid
for a single realization of a point process, Torquato and Stillinger found \cite{ToSt03}:
\begin{eqnarray}\label{volavginterp}
\sigma^2_N(R) = 2^d\phi \left(\frac{R}{D}\right)^d\left[1-2^d\phi\left(\frac{R}{D}\right)^d + \frac{1}{N}\sum_{i\neq j = 1}^N \alpha(r_{ij}; R)\right].
\end{eqnarray} 
The asymptotic coefficient $B_N$ for a hyperuniform point pattern is then given by:
\begin{eqnarray}
B_N &= \lim_{L\rightarrow +\infty} \frac{1}{L}\int_0^L B_N(R) dR,\\
B_N(R) &= \frac{R}{D}\left[1-2^d\phi\left(\frac{R}{D}\right)^d + \frac{1}{N}\sum_{i\neq j = 1}^N \alpha(r_{ij}; R)\right].
\end{eqnarray}
These results imply that the asymptotic coefficient $B_N$ obtained in (\ref{Bn}) involves an average over small-scale fluctuations in the number variance with length
scale on the order of the mean separation between points \cite{ToSt03}.

There is another reformulation of the number variance problem \cite{KeRa53, ToSt03} in terms of the Epstein zeta function that is 
applicable for Bravais lattices.  
By recognizing that the number of points in an observation window for a Bravais lattice is a periodic function of the centroid of the window,
one can establish the following expression for the asymptotic coefficient $\phi^{1/d} B_N$:
\begin{eqnarray}\label{epstein}
\phi^{1/d} B_N = \frac{\pi^{(d-1)/2} 2^{d-1} \left[\Gamma(1+d/2)\right]^{1-1/d}}{v_C^{1+1/d}} \sum_{\mathbf{q}\neq\mathbf{0}} \frac{1}{\Vert\mathbf{q}\Vert^{d+1}},
\end{eqnarray}
where $v_C$ is the volume of the fundamental cell and $\mathbf{q}$ denotes a vector of the dual lattice.  The Epstein zeta function for a lattice is given by \cite{SaSt06}:
\begin{eqnarray}\label{epformula}
Z_{\Lambda}(s) = \sum_{\mathbf{p}\neq\mathbf{0}} \Vert \mathbf{p}\Vert^{-2s} \qquad (\Re s> d/2),
\end{eqnarray}
where $\mathbf{p}$ is a vector of the lattice $\Lambda$.  It is clear from (\ref{epstein}) that the dual of the lattice that minimizes the Epstein zeta function among all lattices 
will minimize the asymptotic coefficient for the number variance among lattices.  We remark that minimization of the Epstein zeta function is equivalent to the problem of 
finding the classical ground state of an inverse power-law pair potential $q^{-(d+1)}$ in dimension $d$, which is completely monotonic.  Torquato and Stillinger have provided certain duality relations \cite{ToSt08} 
that establish rigorous upper bounds on the energies of such ground states and help to identify energy-minimizing lattices.  
It is known in two dimensions that the triangular lattice minimizes the Epstein zeta function \cite{Ra53, Ca59, Di64, En64, SaSt06} among all lattices; in three dimensions
the FCC is at least a local minimum among lattices \cite{En64b, SaSt06}, and for dimensions 4, 8, and 24 the densest known lattice packings are local minima \cite{SaSt06}.  However,
it is almost certainly not true in higher dimensions that the minimizers of the Epstein zeta function are lattice structures since the densest packings are likely 
to be nonperiodic \cite{ToSt06}.  

\section{Local volume fraction fluctuations in two-phase heterogeneous media}

\subsection{Definition of the local volume fraction}

We consider a two-phase heterogeneous random medium, which we now define to be a domain of space $\mathcal{V} \subseteq \mathbb{R}^d$ of volume $V \leq +\infty$ 
that is composed of two regions:
the phase 1 region $\mathcal{V}_1$ of volume fraction $\phi_1$ and the phase 2 region $\mathcal{V}_2$ of volume fraction $\phi_2$ \cite{LuTo90}.
In general $\mathcal{V} = \mathcal{V}(\omega)$ for some realization $\omega$ of an underlying probability space.
The statistical properties of each phase of the system are specified by the countably infinite set of \emph{$n$-point probability functions} $S_n^{(i)}$, which are defined by
\cite{St95, ToSt82, ToSt83, torquato2002rhm}:
\begin{eqnarray}\label{Sndef}
S_n^{(i)}(\mathbf{x}_1, \ldots, \mathbf{x}_n) = \left\langle\prod_{i=1}^n I^{(i)}(\mathbf{x}_i)\right\rangle,
\end{eqnarray}
where $I^{(i)}$ is the indicator function for phase $i$.  If the medium is statistically homogeneous, the volume fraction of phase $i$ is given by:
\begin{eqnarray}
\phi_i = S_1^{(i)}(\mathbf{x}) = \langle I^{(i)}(\mathbf{x})\rangle,
\end{eqnarray}
and this quantity is fixed on the global scale of the microstructure.  However, the volume fraction fluctuates on a local scale determined by an observation 
window $\Omega \subset \mathcal{V}$ of arbitrary geometry as in the case of the local number density.  

Specifically, we define the \emph{local volume fraction} $\tau_i(\mathbf{x})$ of phase $i$ according to:
\begin{eqnarray}\label{one}
\tau_i(\mathbf{x}; R) = \frac{1}{v(R)}\int I^{(i)}(\mathbf{z}) w(\mathbf{z}-\mathbf{x}; R) d\mathbf{z},
\end{eqnarray}
where $v(R)$ is the volume of the observation window and $w$ is the corresponding indicator function.
Using this definition, the variance $\sigma_{\tau}^2$ in the local volume fraction is given by:
\begin{eqnarray}\label{three}
\sigma^2_{\tau_i} = \langle\tau_i^2\rangle - \phi_i^2,
\end{eqnarray}
which leads to the previously mentioned result:
\begin{eqnarray}\label{four}
\sigma_{\tau_i}^2 = \frac{1}{v(R)}\int_{\mathbb{R}^d} S_2(\mathbf{r}) \alpha(\mathbf{r}; R) d\mathbf{r} - \phi_i^2,
\end{eqnarray}
where $S_2$ is the two-point probability function (for phase $i$) and $\alpha$ is the scaled intersection volume.
Note that:
\begin{eqnarray}\label{six}
\frac{1}{v(R)}\int_{\mathbb{R}^d} \alpha(\mathbf{r}; R) d\mathbf{r} = 1;
\end{eqnarray}
therefore, we may express (\ref{four}) in the form:
\begin{eqnarray}\label{seven}
\sigma_{\tau_i}^2 = \frac{1}{v(R)}\int_{\mathbb{R}^d} \left[S_2(\mathbf{r}) - \phi_i^2\right] \alpha(\mathbf{r}; R) d\mathbf{r},
\end{eqnarray}
or, defining the autocovariance function $\chi(\mathbf{r}) = S_2(\mathbf{r}) - \phi_i^2$:
\begin{eqnarray}\label{eight}
\sigma_{\tau_i}^2 = \frac{1}{v(R)} \int_{\mathbb{R}^d} \chi(\mathbf{r}) \alpha(\mathbf{r}; R) d\mathbf{r}.
\end{eqnarray}
We remark that $\chi(\mathbf{r}) \rightarrow \phi_i(1-\phi_i)$ as $\Vert\mathbf{r}\Vert\rightarrow 0$, 
and $\chi(\mathbf{r})\rightarrow 0$ as $\Vert\mathbf{r}\Vert \rightarrow +\infty$ in the absence of 
long-range order.  Thus, for such systems it is true that $\chi\in L^2(\mathbb{R}^d)$, 
and its Fourier transform is well-defined; the same is true for $\alpha(\mathbf{r}; R)$.  


\subsection{Asymptotic behavior of the variance in the local volume fraction}

For a homogeneous and isotropic system with a spherical observation window of radius $R$, the result (\ref{eight}) implies:
\begin{eqnarray}\label{twelve}
\sigma_{\tau}^2 = \frac{1}{v(R)}\int_0^{+\infty} s(r) \chi(r) \alpha(r; R) dr, 
\end{eqnarray}
where $s(r)$ is the surface area of a $d$-dimensional sphere and $v(r)$ is the corresponding volume.  For simplicity we drop the subscript $i$ on the 
local volume fraction with the disclaimer that our results apply for some specified reference phase in the heterogeneous system.
Substitution of the expansion (\ref{series}) for the scaled intersection volume $\alpha(r; R)$ into (\ref{twelve}) implies:
\begin{eqnarray}
&\sigma_{\tau}^2 = \frac{1}{v(R)}\left[\int_0^{2R} s(r)\chi(r)dr - \frac{\kappa(d)}{R}\int_0^{2R} rs(r)\chi(r) dr + o(r/R)\right]\label{fourteen}\\
\Rightarrow &\sigma_{\tau}^2 = \tilde{A}_{\tau}\left(\frac{1}{R^d}\right)+\tilde{B}_{\tau}\left(\frac{1}{R^{d+1}}\right) + o\left(\frac{1}{R^{d+1}}\right)\qquad (R\rightarrow +\infty)\label{fifteen},
\end{eqnarray}
where:
\begin{eqnarray}
\tilde{A}_{\tau} &= \frac{1}{v(1)}\int_{\mathbb{R}^d} \chi(\mathbf{r}) d\mathbf{r}\label{sixteen}\\
\tilde{B}_{\tau} &= -\frac{\kappa(d)}{v(1)}\int_{\mathbb{R}^d} \Vert \mathbf{r}\Vert \chi(\mathbf{r}) d\mathbf{r}\label{seventeen}.
\end{eqnarray}

It is convenient as with the number variance to introduce a length scale $D$, which can, for example, represent the diameter of a spherical 
inclusion in the heterogeneous medium.  The results in (\ref{fifteen})-(\ref{seventeen}) then take the forms:
\begin{eqnarray}
\sigma^2_{\tau} &= \frac{\rho}{2^d\phi}\left\{A_{\tau} \left(\frac{D}{R}\right)^d + B_{\tau} \left(\frac{D}{R}\right)^{d+1} 
+ o\left[\left(\frac{D}{R}\right)^{d+1}\right]\right\}\label{tauasymp}\\
A_{\tau} &= \int_{\mathbb{R}^d} \chi(\mathbf{r}) d\mathbf{r}\label{atau}\\
B_{\tau} &= -\frac{\kappa(d)}{D} \int_{\mathbb{R}^d} \Vert \mathbf{r}\Vert \chi(\mathbf{r}) d\mathbf{r}\label{btau}.
\end{eqnarray}
The coefficients $A_{\tau}$ and $B_{\tau}$ in (\ref{atau}) and (\ref{btau}) control the asymptotic scaling of the fluctuations in the local volume fraction.  
We note that:
\begin{eqnarray}\label{eighteen}
A_{\tau} = \lim_{\Vert\mathbf{k}\Vert\rightarrow 0}\hat{\chi}(\mathbf{k}),
\end{eqnarray}
where $\hat{\chi}$ is the Fourier transform of $\chi$.  It then follows that $\sigma_{\tau}^2 \sim R^{-(d+1)}$ as $R\rightarrow +\infty$ for those systems such that: 
\begin{eqnarray}\label{nineteen}
\lim_{\Vert\mathbf{k}\Vert\rightarrow 0}\hat{\chi}(\mathbf{k}) = 0.
\end{eqnarray}
This behavior is for random heterogeneous media the equivalent to hyperuniformity in the description of local number density fluctuations in 
stochastic point processes.  

\section{An estimate of $B_{\tau}$ using $B_N$ for microstructures of spherical inclusions}

It is possible to derive a bound for the asymptotic coefficient $B_{\tau}$ governing fluctuations in the local volume fraction in terms of the 
corresponding coefficient $B_N$ for the local number density
in the case of microstructures of spherical inclusions of radius $\lambda = D/2$. 
We first consider a system of impenetrable spheres and then generalize our results to the penetrable case.
Note that this setting is quite general, incorporating lattice and ordered packing 
structures below the maximal packing fraction in addition to equilibrium and nonequilibrium distributions of hard spheres.     
To establish this connection, we first note that the series representation of $S_2$ for the matrix phase external to any inclusions
is given in terms of the $n$-particle correlation functions $g_n$ by 
\cite{ToSt82, ToSt83, ToSe90, LuTo90, torquato2002rhm}:
\begin{eqnarray}\label{S2series}
S_2(\mathbf{x}, \mathbf{y}) &= 1+\sum_{k=1}^{+\infty} \frac{(-1)^k \rho^k}{k!}\int g_k(\mathbf{r}^k)\times\nonumber\\ 
&\prod_{j=1}^{k} \left[m(\mathbf{x}-\mathbf{r}_j)+m(\mathbf{y}-\mathbf{r}_j)
-m(\mathbf{x}-\mathbf{r}_j)m(\mathbf{y}-\mathbf{r}_j)\right] d\mathbf{r}_j,
\end{eqnarray}
where $m$ is a particle indicator function for an inclusion in phase $i$.  For the special case of inclusions of 
statistically homogeneous disjoint impenetrable spheres, (\ref{S2series})
simplifies according to \cite{LuTo90}:
\begin{eqnarray}\label{S2impen}
S_2(r) = 1-2\phi+\rho v_{\mbox{int}}(r; \lambda) + \rho^2(g_2*v_{\mbox{int}})(r; \lambda),
\end{eqnarray}
where the last term is the convolution of the pair correlation function with the spherical intersection volume
and $\phi$ is the volume fraction of inclusions.  The result in (\ref{S2impen}) immediately implies that the autocovariance function $\chi$ for this 
system is:
\begin{eqnarray}\label{autocorrelHS}
\chi(r) = \rho v_{\mbox{int}}(r; \lambda) + \rho^2 (h*v_{\mbox{int}})(r; \lambda),
\end{eqnarray}
and this equation is valid for either phase of the heterogeneous medium.  It now follows from (\ref{atau}) that the coefficient $A_{\tau}$ is given by:
\begin{eqnarray}
A_{\tau} &= \rho \left[v(\lambda)\right]^2 + \left[\rho v(\lambda)\right]^2 \int_{\mathbb{R}^d} h(\mathbf{r}) d\mathbf{r}\\
 &= \rho\left[v(\lambda)\right]^2\left[1+\rho \int_{\mathbb{R}^d} h(\mathbf{r}) d\mathbf{r}\right]\\
&= \phi^2 A_N/\rho\label{atan}.
\end{eqnarray}
One can see that a hyperuniform point pattern derived from a system of impenetrable spheres 
generates a hyperuniform heterogeneous medium with respect to fluctuations in the local volume fraction.  It is also
clear that the leading-order asymptotic coefficient for the variance in the local volume fraction can be determined completely from the knowledge of $A_N$,
a more accessible quantity since it does not require the explicit calculation of the autocovariance function $\chi$.  

The calculation of $B_{\tau}$ is generally nontrivial, even for a medium of impenetrable spheres.  Using an argument similar to the one for $A_{\tau}$,
one can establish the following exact relation between $B_{\tau}$ and $B_N$:
\begin{eqnarray}\label{btbn}
B_{\tau} = \left(\frac{B_N}{\rho}\right)\frac{\int_{\mathbb{R}^d} \Vert\mathbf{r}\Vert \chi(\mathbf{r}) d\mathbf{r}}{\int_{\mathbb{R}^d} \Vert \mathbf{r}\Vert
h(\mathbf{r}) d\mathbf{r}};
\end{eqnarray}
there is no simple factorization of the $d$-th moment of the convolution $(h*v_{\mbox{int}})$ for which (\ref{btbn}) simplifies.  Therefore, the coefficient 
$B_{\tau}$ depends explicitly on the two-point information of the heterogeneous system.  Nevertheless, one can establish a rigorous upper bound on $B_{\tau}$ in 
terms of $B_N$.  Toward this end, we first note that the local volume fraction $\tau$ of $N$ impenetrable spheres centered at 
$\left\{\mathbf{r}_i\right\}_{i=1}^N$ with respect to a spherical observation window of radius $R$ is given by:
\begin{eqnarray}\label{impspheretau}
\tau = \frac{1}{v(R)}\sum_{i=1}^N v_{\mbox{int}}(\Vert \mathbf{r}_i - \mathbf{x}_0\Vert; R, \lambda),
\end{eqnarray}
where $\mathbf{x}_0$ is the center of the observation window and $v_{\mbox{int}}(\Vert\mathbf{r}\Vert; R, \lambda)$ is the intersection volume between a sphere of 
radius $\lambda$ and another sphere of radius $R$ given a separation $\Vert\mathbf{r}\Vert$ between their centers.  
If the underlying probability distribution for the sphere centers is given by $P_N(\mathbf{r}^N)$, 
then we immediately recover the result:
\begin{eqnarray}
\langle\tau\rangle &= \frac{\rho}{v(R)} \int_{\mathbb{R}^d} v_{\mbox{int}}(\Vert\mathbf{r}\Vert; R, \lambda) d\mathbf{r}\\
&= \rho v(\lambda) = \phi\label{avtau},
\end{eqnarray}
where we have assumed statistical homogeneity and used the identity:
\begin{eqnarray}
\rho = N\int P_N(\mathbf{r}, \mathbf{r}^{N-1}) d\mathbf{r}^{N-1}.
\end{eqnarray}
Using (\ref{impspheretau}), we may also write:
\begin{eqnarray}
\tau^2 &= \sum_{i=1}^N \left[\frac{v_{\mbox{int}}(\Vert\mathbf{r}_i - \mathbf{x}_0\Vert; R, \lambda)}{v(R)}\right]^2 \nonumber\\
&+ \left[\frac{1}{v(R)}\right]^2 
\sum_{i\neq j = 1}^N v_{\mbox{int}}(\Vert\mathbf{r}_i - \mathbf{x}_0 \Vert; R, \lambda) v_{\mbox{int}}(\Vert \mathbf{r}_j - \mathbf{x}_0\Vert; R, \lambda),
\end{eqnarray}
which implies:
\begin{eqnarray}\label{tau2}
\fl \langle\tau^2\rangle = \frac{\rho}{\left[v(R)\right]^2}
\int_{\mathbb{R}^d} v_{\mbox{int}}^2(\Vert\mathbf{r}\Vert; R, \lambda)d\mathbf{r} + 
\left[\frac{\rho}{v(R)}\right]^2 \int_{\mathbb{R}^{d}} g_2(\Vert\mathbf{r}\Vert) (v_{\mbox{int}}* v_{\mbox{int}})(\Vert\mathbf{r}\Vert) d\mathbf{r},
\end{eqnarray}
where $(v_{\mbox{int}} * v_{\mbox{int}})$ is the self-convolution of the intersection volume.  
The expression in (\ref{tau2}) admits the following upper bound:
\begin{eqnarray}
\langle\tau^2\rangle &\leq \rho \left[\frac{v(\lambda)}{v(R)}\right]^2 v(R+\lambda) + \left[\frac{\rho v(\lambda)}{v(R)}\right]^2 
\int_{\mathbb{R}^d} g_2(\Vert\mathbf{r}\Vert)
v_{\mbox{int}}(\Vert\mathbf{r}\Vert; R+\lambda) d\mathbf{r}\label{ineq}\\
 &= \rho \left[\frac{v(\lambda)}{v(R)}\right]^2 v(R+\lambda)
\left[1+\rho\int_{\mathbb{R}^d} g_2(\Vert\mathbf{r}\Vert) \alpha(\Vert \mathbf{r}\Vert; R+\lambda) d\mathbf{r}\right]\label{integral}\\
&= \left[\frac{v(\lambda)}{v(R)}\right]^2 \langle N^2(R+\lambda)\rangle\label{tauN},
\end{eqnarray}
where $\alpha(r; R+\lambda) = v_{\mbox{int}}(r; R+\lambda)/v(R+\lambda)$.
The result (\ref{ineq}) uses the inequality $v_{\mbox{int}}(\Vert\mathbf{r}\Vert; R, \lambda) \leq v(\lambda) w(\Vert\mathbf{r}\Vert; R+\lambda),$
where $w(r; R+\lambda)$ is the indicator function for a sphere of radius $R+\lambda$.  
Combining (\ref{avtau}) and (\ref{tauN}), we find:
\begin{eqnarray}
\sigma^2_{\tau}(R) &\leq \left[\frac{\phi}{\rho v(R)}\right]^2 \langle N^2(R+\lambda)\rangle - \left[\frac{\phi}{\rho v(R+\lambda)}\right]^2 \langle N(R+\lambda)\rangle^2\\
  &= \left[\frac{\phi}{\rho v(R)}\right]^2 \left\{\langle N^2(R+\lambda)\rangle - \left[\frac{v(R)}{v(R+\lambda)}\right]^2 \langle N(R+\lambda)\rangle^2\right\};
\end{eqnarray}
in the asymptotic limit $\lambda/R  \ll 1$ we find :
\begin{eqnarray}\label{NtauUL}
\sigma^2_{\tau}(R) \leq \left[\frac{\phi}{\rho v(R)}\right]^2 \sigma^2_N(R) \qquad (R\rightarrow +\infty).
\end{eqnarray}
Substituting the asymptotic expansions (\ref{numasymp}) and (\ref{tauasymp}) into (\ref{NtauUL}) gives an upper bound that is exact to $\mathcal{O}\left[(D/R)^{d}\right]$
according to (\ref{atan}).  It is therefore true that:
\begin{eqnarray}
B_{\tau} &\leq \frac{\phi^2 B_N}{\rho}\label{BNtau}.
\end{eqnarray}
The relation (\ref{BNtau}) is generally not an equality for a given 
heterogeneous system, and we provide a specific example in Section 6 where the strict inequality holds.

We remark that in situations where the microstructure of the heterogeneous medium consists of overlapping spheres, the variance in the local volume fraction 
is generally expected to be greater than the corresponding fluctuations for impenetrable inclusions at a fixed volume fraction $\phi$ \cite{LuTo90, QuTo97, torquato2002rhm}.
This behavior arises since the excluded volume induced by an impenetrable shell formally induces a greater degree of uniformity in the underlying point process
generated by the sphere centers, thereby reducing the variance in the local volume fraction.  However, at a \emph{fixed reduced density} $\eta = \rho v(\lambda)$,
which is greater than the volume fraction of penetrable spheres,  
the general upper bound 
given in (\ref{NtauUL}) (with the replacement $\phi \rightarrow \eta$) will still hold.
This claim is a direct result of the relation (\ref{impspheretau}), the right-hand side of 
which becomes an upper bound for overlapping spheres at fixed $\rho$.  The remainder of the analysis leading to (\ref{NtauUL}) therefore also works for this case with
the exception that, for nonhyperuniform systems, the bound (\ref{BNtau}) becomes a bound on the leading-order coefficients $A_{\tau}$ and $A_N$\footnote{Careful
replication of the analysis for hard spheres in the penetrable case actually shows $\sigma^2_{\tau} \leq \{\eta/[\rho v(R)]\}^2 \sigma^2_N + \mbox{constant}$; the
constant depends only on the reduced density $\eta$ and the volume fraction $\phi$ and plays no role for large windows.}.  
For hyperuniform point patterns, the identity $A_N = 0$ and the bound (\ref{NtauUL}) immediately imply $A_{\tau} = 0$, even for microstructures of penetrable spheres,
and we then retain the result (\ref{BNtau}) on $B_{\tau}$ and $B_N$.  

\section{Low-dimensional local density fluctuations and order metrics}

Although Torquato and Stillinger have provided some calculations of the asymptotic coefficients $B_N$ for certain periodic and disordered
hyperuniform point patterns \cite{ToSt03}, an extensive classification of hyperuniform systems remains an active research area.  Here we provide 
new calculations of the asymptotic coefficients for selected periodic and nonperiodic hyperuniform point patterns, including the Fibonacci chain, 
the so-called ``tunneled crystals'' introduced elsewhere by Torquato and Stillinger \cite{ToSt07}, a four-coordinate relative of the Kagom\'e lattice, the 4.8.8
Archimedean tiling of the plane, and certain classical disordered ground states in low Euclidean dimensions.

Our major results, including detailed tables of number variance coefficients for 
several hyperuniform point patterns, are collected in subsection 1.  Since the current literature on hyperuniform point processes is disparate, we have provided
details on several periodic, quasiperiodic, and nonperiodic hyperuniform point patterns, including processes introduced in this article and 
previously by other authors, in subsections 2 and 3.  The information in these subsections is meant to supplement the key results of this article
and should provide a brief and self-contained (though almost certainly not complete) overview of the field.  Those readers interested only 
in the results may choose to skim this material.  

\subsection{Number variance coefficients for $d = 1, 2,$ and $3$}

We collect in Tables \ref{d1table}-\ref{tabletwo} the asymptotic coefficients $B_N$ for several hyperuniform point patterns, and the results 
establish a quantitative \emph{order metric} for 
point patterns, meaning that they measure the degree of regularity in the spatial distribution of the points.  The details of accurately calculating $B_N$ for
a given point process have been provided elsewhere \cite{ToSt03}; described below are a number of pertinent results for characterizing the fluctuations in a hyperuniform 
point pattern.  

The result (\ref{Bn}) for the calculation of $B_N$ cannot in general be applied to periodic point patterns because such systems are not statistically homogeneous. 
Using the following radial form of the pair correlation function:
\begin{eqnarray}
g_2(r) = \sum_{k=1}^{+\infty} \frac{Z_k\delta(r-r_k)}{\rho s(r_k)},
\end{eqnarray}
we find that the total correlation function $h$ is nonintegrable.  However, one can obtain a convergent expression for $B_N$ using the following 
interpretation of (\ref{Bn}) \cite{ToSt03}:
\begin{eqnarray}
B_N = \lim_{\beta\rightarrow 0^+} \frac{-\phi \kappa(d)}{D v(D/2)} \int_{\mathbb{R}^d}\Vert\mathbf{r}\Vert \exp\left(-\beta \Vert\mathbf{r}\Vert^2\right) 
h(\mathbf{r}) d\mathbf{r},
\end{eqnarray}
which implies:
\begin{eqnarray}\label{Bnconv}
\fl B_N = \lim_{\beta\rightarrow 0^+} \frac{d}{2\sqrt{\pi} D}\left[\frac{\phi \pi^{d/2}}{v(D/2) \beta^{(d+1)/2}} - \frac{\Gamma(d/2)}{\Gamma[(d+1)/2]}
\sum_{k=1}^{+\infty} Z_k r_k \exp\left(-\beta r_k^2\right)\right].
\end{eqnarray}  
We note that to compare values of $B_N$ among different hyperuniform systems one should use the rescaled coefficient $\phi^{1/d}B_N$; reference to (\ref{Bnconv})
shows that this scaling eliminates the dependence on both the length scale $D$ and the reduced density $\phi$.  For $d = 1, 2, 3$ and $D = 1$, (\ref{Bnconv}) takes the forms:
\begin{eqnarray}
B_N &= \lim_{\beta\rightarrow 0^+} \left[\frac{\phi}{2\beta} - \frac{1}{2}\sum_{k=1}^{+\infty} Z_k r_k \exp\left(-\beta r_k^2\right)\right] \qquad (d=1)\label{Bnd1}\\
B_N &= \lim_{\beta\rightarrow 0^+} \left[\frac{4\phi}{\sqrt{\pi} \beta^{3/2}} - \frac{2}{\pi}\sum_{k=1}^{+\infty} Z_k r_k \exp\left(-\beta r_k^2\right)\right] \qquad (d=2)\\
B_N &= \lim_{\beta\rightarrow 0^+} \left[\frac{9\phi}{\beta^2} - \frac{3}{4}\sum_{k=1}^{+\infty} Z_k r_k \exp\left(-\beta r_k^2\right)\right] \qquad (d=3)\label{Bnd3}.
\end{eqnarray}

The summations in (\ref{Bnd1})-(\ref{Bnd3}) can be easily accomplished using the \emph{theta series} for a periodic point pattern $\Lambda$; it is defined by:
\begin{eqnarray}
\Theta_{\Lambda}(q) = 1+\sum_{k=1}^{+\infty} Z_k q^{r_k^2},
\end{eqnarray}
where $r_k$ is the distance to the $k$-th coordination shell and $Z_k$ is the corresponding coordination number.  This series can usually be generated from 
the simpler functions $\theta_2,$ $\theta_3$, and $\theta_4$, which are defined by \cite{CoSl99}:
\begin{eqnarray}
\theta_2(q) &= 2\sum_{m=0}^{+\infty} q^{(m+1/2)^2}\\
\theta_3(q) &= 1+2\sum_{m=1}^{+\infty} q^{m^2}\\
\theta_4(q) &= 1+2\sum_{m=1}^{+\infty} (-q)^{m^2}.
\end{eqnarray}
These functions have the following intuitive interpretations:  $\theta_2$ and $\theta_3$ assign points to the half-integer and integer positions, respectively, on the line,
and $\theta_2$ assigns points to the even integers and removes points from the odd integers \cite{CoSl99}.  It thus follows that the theta series for the 
integer lattices $\mathbb{Z}^d$ and checkerboard lattices $D_d$ are given respectively by:
\begin{eqnarray}
\Theta_{\mathbb{Z}^d} &= \left[\theta_3(q)\right]^d\\
\Theta_{D_d} &= \frac{1}{2}\left\{\left[\theta_3(q)\right]^d + \left[\theta_4(q)\right]^d\right\}.
\end{eqnarray}
The theta series for most well-known lattices have been tabulated analytically \cite{CoSl99, Sl87}; they are directly related to the quadratic forms 
associated with the lattices.    

\begin{figure}[!tp]
\centering
\includegraphics[width=0.35\textwidth]{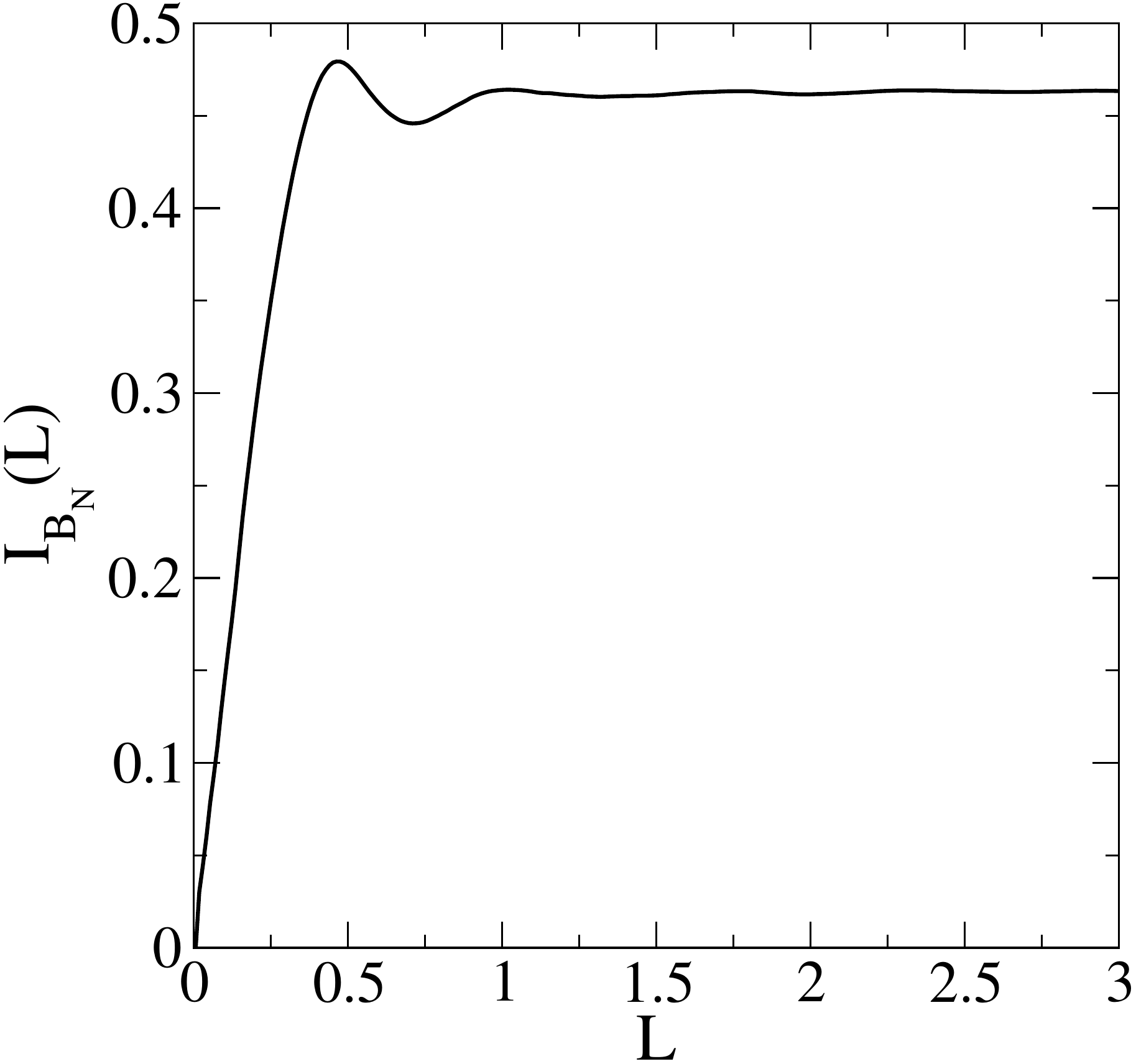}
\caption{Typical fluctuations for a disordered point pattern in the ``running average'' of the asymptotic coefficient $B_N(R)$ from (\ref{Bnsmall}).  
$I_{B_N}(L) = (1/L)\int_{0}^L B_N(R) dR$.
Note that the unit of length (here given by the box size) is $D \approx 23.5$ for this system, meaning that fluctuations have almost completely dissipated upon reaching approximately $4\%$
of the length scale.}\label{runavg}
\end{figure}
With regard to nonperiodic systems, Torquato and Stillinger have developed a volume-average interpretation of the characterization of hyperuniform point patterns 
\cite{ToSt03}; namely, the following relations hold for any single realization of a statistically homogeneous ergodic point process:
\begin{eqnarray}
&B_N(R) = \frac{R}{D}\left[1-2^d\phi \left(\frac{R}{D}\right)^d + \frac{1}{N}\sum_{i\neq j=1}^N \alpha(r_{ij}; R)\right]\label{Bnsmall}\\
&B_N = \lim_{L\rightarrow +\infty} \frac{1}{L} \int_0^L B_N(R) dR\label{Bnav}.
\end{eqnarray}
Note again that using the rescaled coefficient $\phi^{1/d} B_N$ eliminates the dependence of the asymptotic coefficient on $D$ and $\phi$ according 
to (\ref{Bnsmall}); the length scale for the system is then determined by the reciprocal number density $\rho^{-1/d}$, and, making use of 
reduced units, it follows that once the problem has been solved for unit number density, it has been solved for all number densities.
The result (\ref{Bnsmall}) captures small-scale fluctuations with respect to $R$ with a wavelength on the order of the mean separation between points; these 
fluctuations are superimposed on the large-scale variations with respect to $R$ \cite{ToSt03}. These relations are useful in characterizing fluctuations in the 
local number density for disordered point patterns, where only limited information about the thermodynamic limit may be available.  We remark, however, that 
any small-scale fluctuations in $R$ typically only appear for $\sim 3$ nearest-neighbor distances and rapidly vanish; as a result, one can obtain quite reliable 
results for $B_N$ given a realization of a point process with $\sim 100$ particles.  Ensemble averaging improves these calculations, but large numbers of configurations
are not always necessary to characterize a given point process.  As few as $5$ configurations is empirically sufficient to determine $B_N$ using (\ref{Bnsmall}) and 
(\ref{Bnav}).  Figure \ref{runavg} demonstrates the dissipation in the small-$R$ fluctuations of $B_N(R)$.    

\Table{\label{d1table} Asymptotic coefficients and order metrics for selected $d = 1$ hyperuniform systems. Note that $\overline{\Lambda} = 2^d \phi B_N$ is
simply a rescaling of the asymptotic coefficient as adopted in \cite{ToSt03}.}
\br
System & $\phi B_N$ & $\overline{\Lambda}$ & $\psi_N$\\
\mr
$\mathbb{Z}$ & 0.08333 & 0.16667 & 1.00000\\
step+delta-function $g_2$ & 0.09375 & 0.18750 & 0.88889\\
Fibonacci chain & 0.10055 & 0.20110 & 0.82878\\
step-function $g_2$ & 0.12500 & 0.25 & 0.66667\\
two-scale lattice & 0.14583 & 0.29167 & 0.57143\\
lattice gas & 0.16667 & 0.33333 & 0.50000\\
\br
\endTable
\Table{\label{tableone}  Asymptotic coefficients and order metrics for selected $d = 2$ hyperuniform systems.  Note that $\overline{\Lambda} = 2^d \phi B_N$ is
simply a rescaling of the asymptotic coefficient as adopted in \cite{ToSt03}.}
\br
System & $\phi^{1/2} B_N$ & $\overline{\Lambda}/\phi^{1/2}$ & $\psi_N$\\
\mr
$A_2$ (triangular) & 0.12709 & 0.50835  & 1.00000\\
$\mathbb{Z}^2$ & 0.12910 & 0.51640 & 0.98443\\
disordered GS; $\gamma = 0.496$ & 0.13100 & 0.52400 & 0.97015\\
disordered GS; $\gamma = 0.402$ & 0.13454 & 0.53816 & 0.94463\\
honeycomb & 0.14176 & 0.56703  & 0.89652\\
Kagom\' e & 0.14675 & 0.58699 & 0.86603\\
octagonal tiling & 0.14892 & 0.59567 & 0.85341\\
step+delta-function $g_2$ & 0.15005 & 0.60021 & 0.84698\\
Penrose tiling & 0.15013 & 0.60052 & 0.84652\\
4-coordinate Kagom\' e relative & 0.15248 & 0.60990 & 0.83349\\
disordered GS; $\gamma = 0.302$ & 0.15314 & 0.61254 & 0.82989\\
$4.8.8$ tessellation & 0.17497 & 0.69987 & 0.72635\\
step-function $g_2$ & 0.21221 & 0.84883 & 0.59889\\
One-component plasma & 0.28210 & 1.12838 & 0.45051\\
\br
\endTable
\Table{\label{tabletwo}  Asymptotic coefficients and order metrics for selected $d = 3$ hyperuniform systems. Note that $\overline{\Lambda} = 2^d \phi B_N$ is
simply a rescaling of the asymptotic coefficient as adopted in \cite{ToSt03}.}
\br
System & $\phi^{1/3} B_N$ & $\overline{\Lambda}/\phi^{2/3}$ & $\psi_N$\\
\mr
$D_3^*$ (BCC) & 0.15560 & 1.24476 & 1.00000 \\
$D_3$ (FCC) & 0.15569 & 1.24552 & 0.99942 \\
HCP & 0.15571 & 1.24569 & 0.99929 \\
$\mathbb{Z}^3$ & 0.16115 & 1.28920 & 0.96556 \\
disordered GS; $\gamma = 0.43$ & 0.16217 & 1.29737 & 0.95949 \\
$D_3^+$ (diamond) & 0.17737 & 1.41892 & 0.87726 \\
tunneled FCC & 0.17760 & 1.42080 & 0.87613 \\
w\" urzite & 0.17773 & 1.42184 & 0.87549 \\
tunneled HCP & 0.17856 & 1.42845 & 0.87142 \\
damped-oscillating $g_2$ & 0.18105 & 1.44837 & 0.85943 \\
step+delta-function $g_2$ & 0.19086 & 1.52686 & 0.81526 \\
step-function $g_2$ & 0.28125 & 2.25000 & 0.55324 \\
\br
\endTable

It is an open problem in the statistical physics of many-particle systems to construct simple scalar order metrics that quantify the degree of 
``randomness'' in a point pattern.  Such metrics are essential in defining the maximally random jammed state as discussed above and
represent a fundamental way of characterizing point patterns in any dimension.  One notices from the results in Tables \ref{d1table}-\ref{tabletwo}
that lattice structures possess smaller asymptotic number variance coefficients compared to random point patterns; in other words, as a point pattern
more closely resembles a Poisson point process, the number variance grows more quickly with the size of an observation window for large windows.  
This observation suggests that by normalizing the asymptotic coefficient $\phi^{1/d} B_N$ by the corresponding result for the 
variance-minimizing structure, one can obtain a scalar quantity between 0 and 1 characterizing the extent of \emph{global} order in the system.  Specifically, we define the 
following order metric for a point pattern $\Lambda$ relative to the variance-minimizing structure $\Lambda_{\mbox{min}}$ for a given dimension:
\begin{eqnarray}\label{metric}
\psi^{(\Lambda)}_N \equiv \frac{\phi_{\Lambda_{\mbox{min}}}^{1/d} B^{(\Lambda_{\mbox{min}})}_N}{\phi_{\Lambda}^{1/d} B^{(\Lambda)}_N}.
\end{eqnarray}

We remark that the metric proposed in (\ref{metric}) is only one of many possible choices for characterizing order in a many-particle system.  
Since complete information about such a system is generally unavailable, one must almost always settle for \emph{reduced information} about 
a system as contained in $\psi^{(\Lambda)}_N$, which depends implicitly on the two-particle information of the point process.  One advantageous property of 
(\ref{metric}) is that the choice of the reference structure is \emph{essentially} not arbitrary since it is quantitatively well-defined 
by the variance-minimizing structure.  It is possible (and in fact likely in higher dimensions) that such a structure will be degenerate,
meaning that it will share the same two-point information with a number of other systems; however, this degeneracy does not affect the 
quantitative values of $\psi^{(\Lambda)}_N$ for all other structures once the minimum value of $\phi^{1/d} B_N$ has been identified.  
This behavior is in contrast to other orientational and translational metrics \cite{StNeRo83, ToTrDe00}, which 
in three dimensions arbitrarily define the FCC lattice as the standard of perfect spatial ordering.  Additionally, the fact that 
the order metric (\ref{metric}) classifies a quasicrystalline structure such as the Fibonacci chain or the Penrose tiling as being
between perfect crystalline point patterns and correlated disordered systems is noteworthy because it is consistent with the 
notion that quasicrystals possess features of both of these types of systems.  In other words, it is clear from Tables \ref{d1table}-\ref{tabletwo}
that quasicrystals possess less order than any Bravais lattice, but there do exist ``structured'' disordered point patterns and periodic nonlattices
with number variance coefficients of roughly the same magnitudes as quasicrystals.  It is this observation which in part suggests the usefulness of 
hyperuniformity in understanding the degree of order within a point pattern; namely, we are able to identify a restricted set of point processes
with an quantitative order metric that matches our intuition of the relative extent of (dis)order within a system.

\subsection{Known hyperuniform systems}
\begin{figure}[!tp]
\centering
\includegraphics[width=0.5\textwidth]{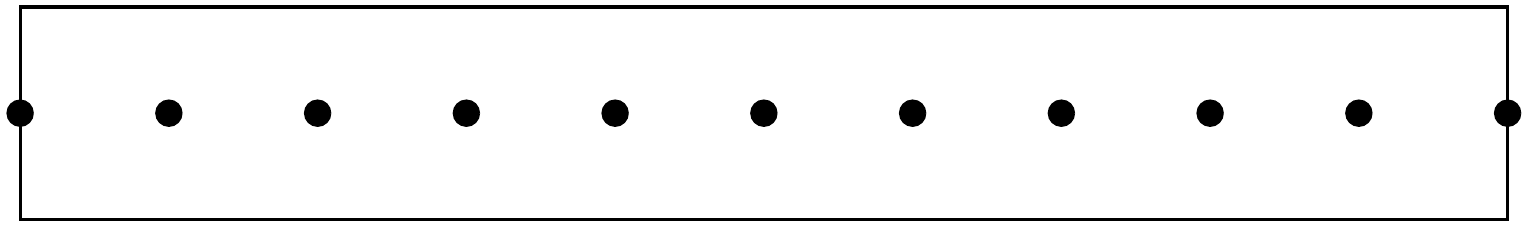}
\includegraphics[width=0.5\textwidth]{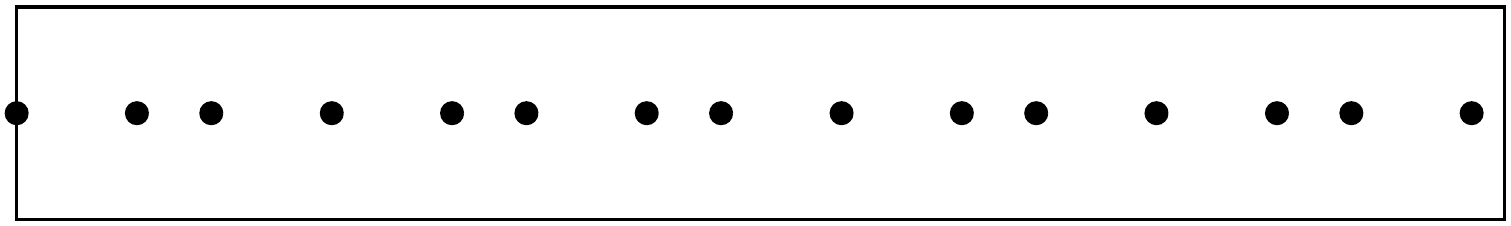}
\includegraphics[width=0.5\textwidth]{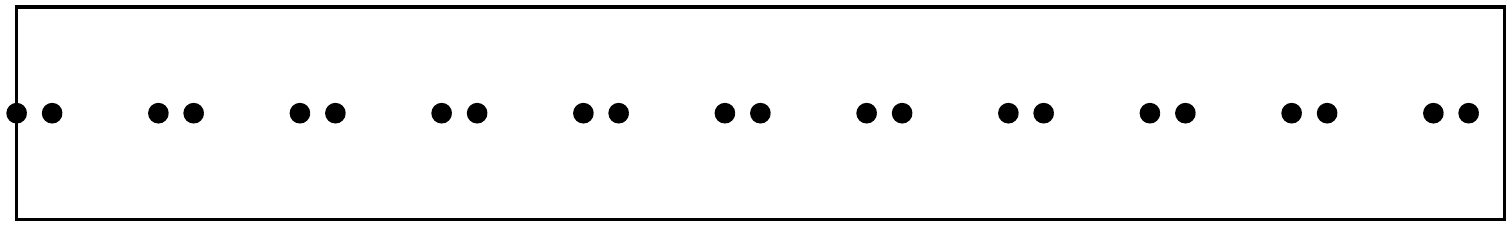}
\caption{\emph{Top:}  The $d = 1$ integer lattice $\mathbb{Z}$.  \emph{Center:}  The Fibonacci chain, a prototypical quasicrystal.  \emph{Bottom:}  A two-scale
 $d = 1$ point pattern.}\label{d1latt}
\end{figure}

\begin{figure}[!tp]
\centering
\includegraphics[width=0.3\textwidth]{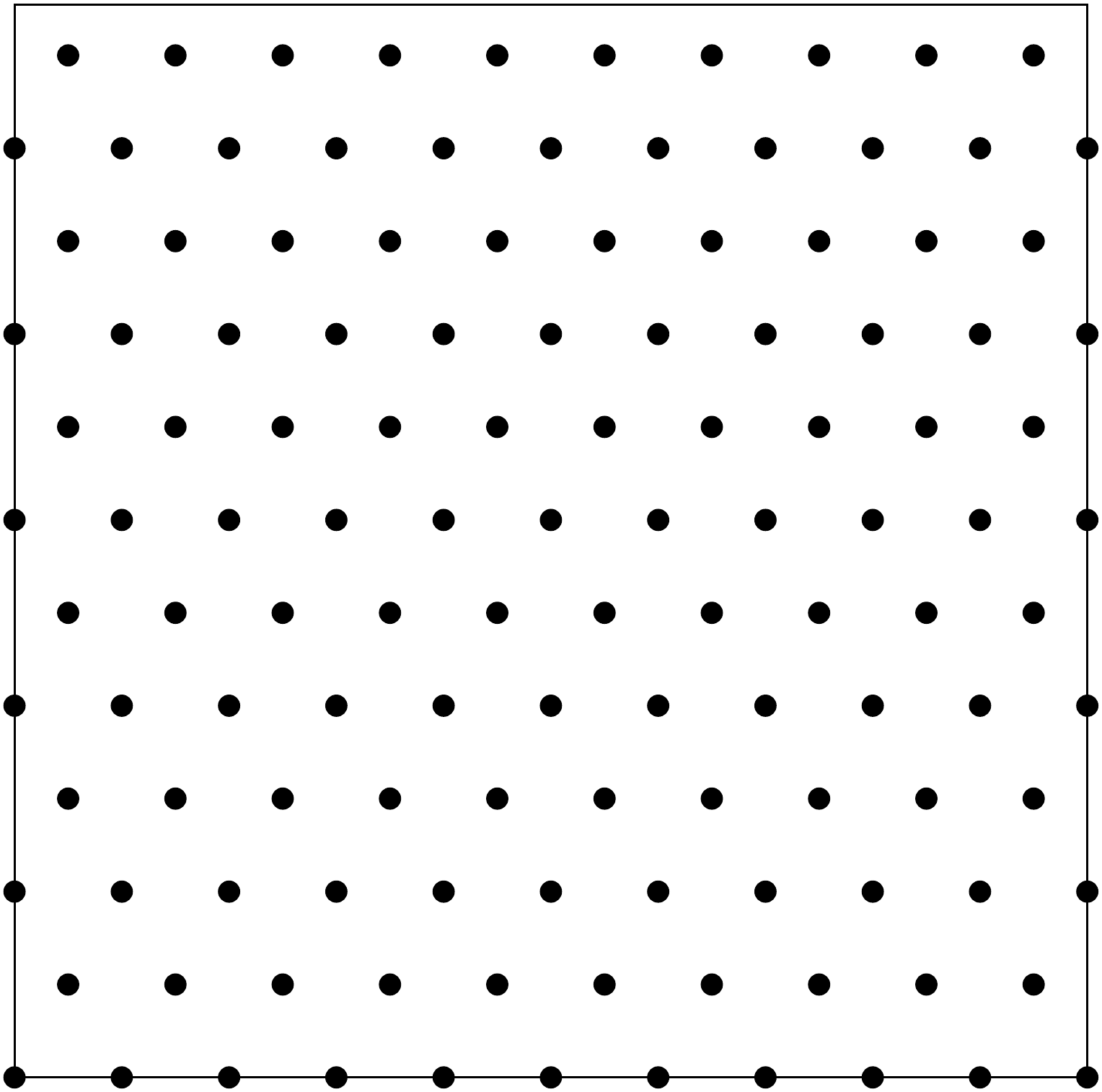}\hspace{0.03\textwidth}
\includegraphics[width=0.3\textwidth]{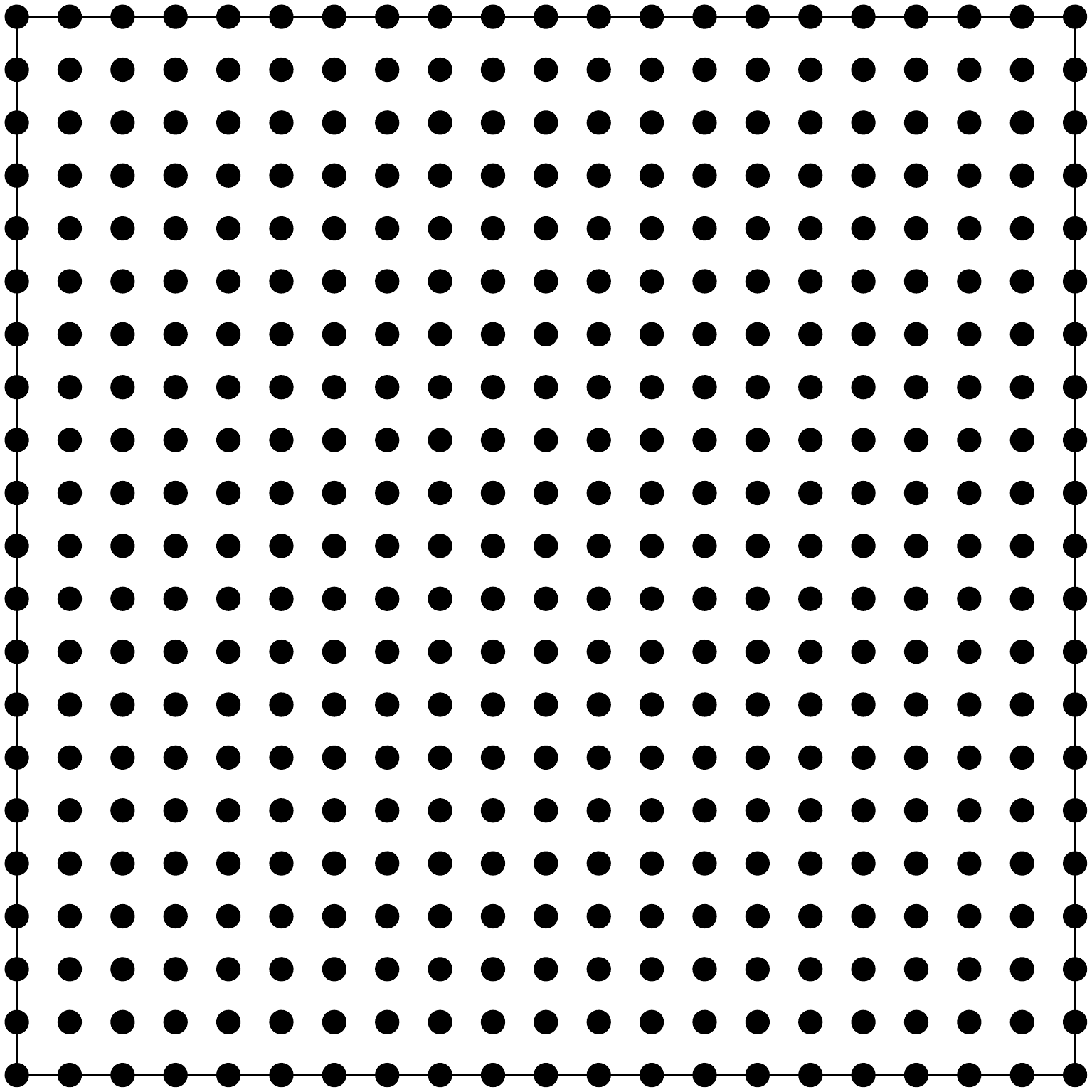}\hspace{0.03\textwidth}
\includegraphics[width=0.3\textwidth]{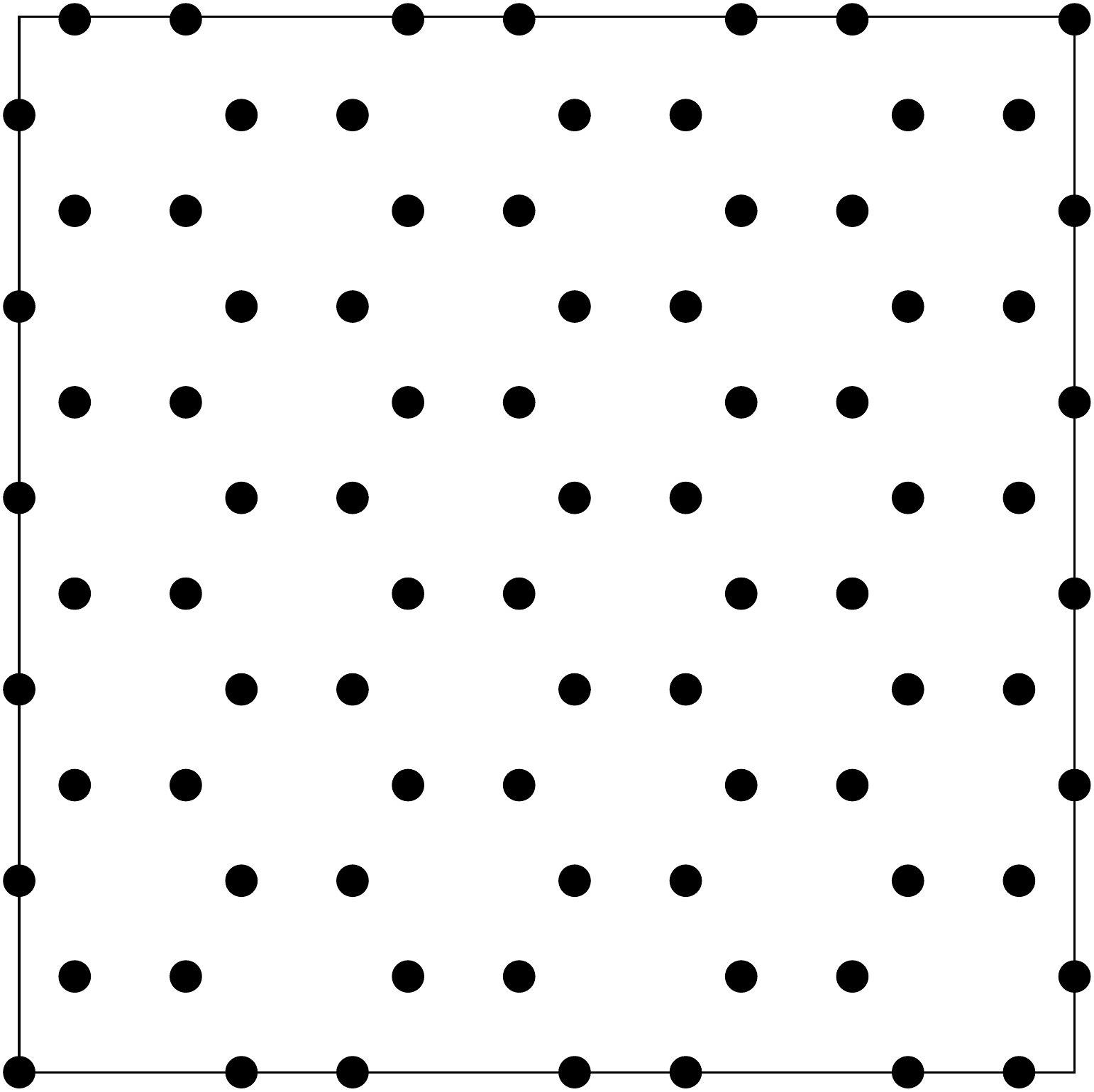}
\caption{\emph{Left:}  The triangular lattice $A_2$.  \emph{Center:}  The square lattice $\mathbb{Z}^2$.  \emph{Right}:  The honeycomb periodic point pattern.}\label{d2latt1}
\end{figure}

\begin{figure}[!tp]
\centering
\includegraphics[width=0.35\textwidth]{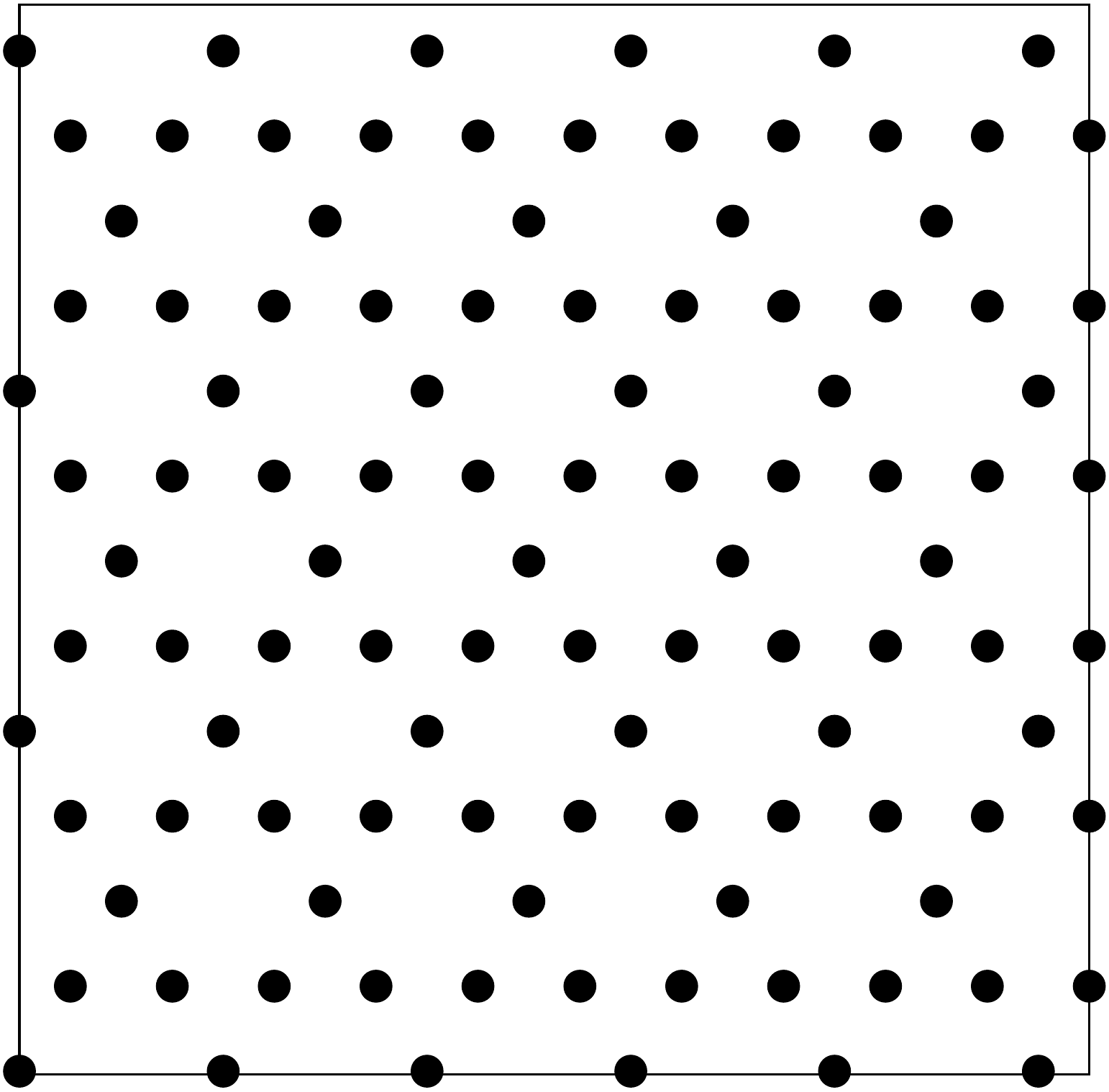}\hspace{0.05\textwidth}
\includegraphics[width=0.35\textwidth]{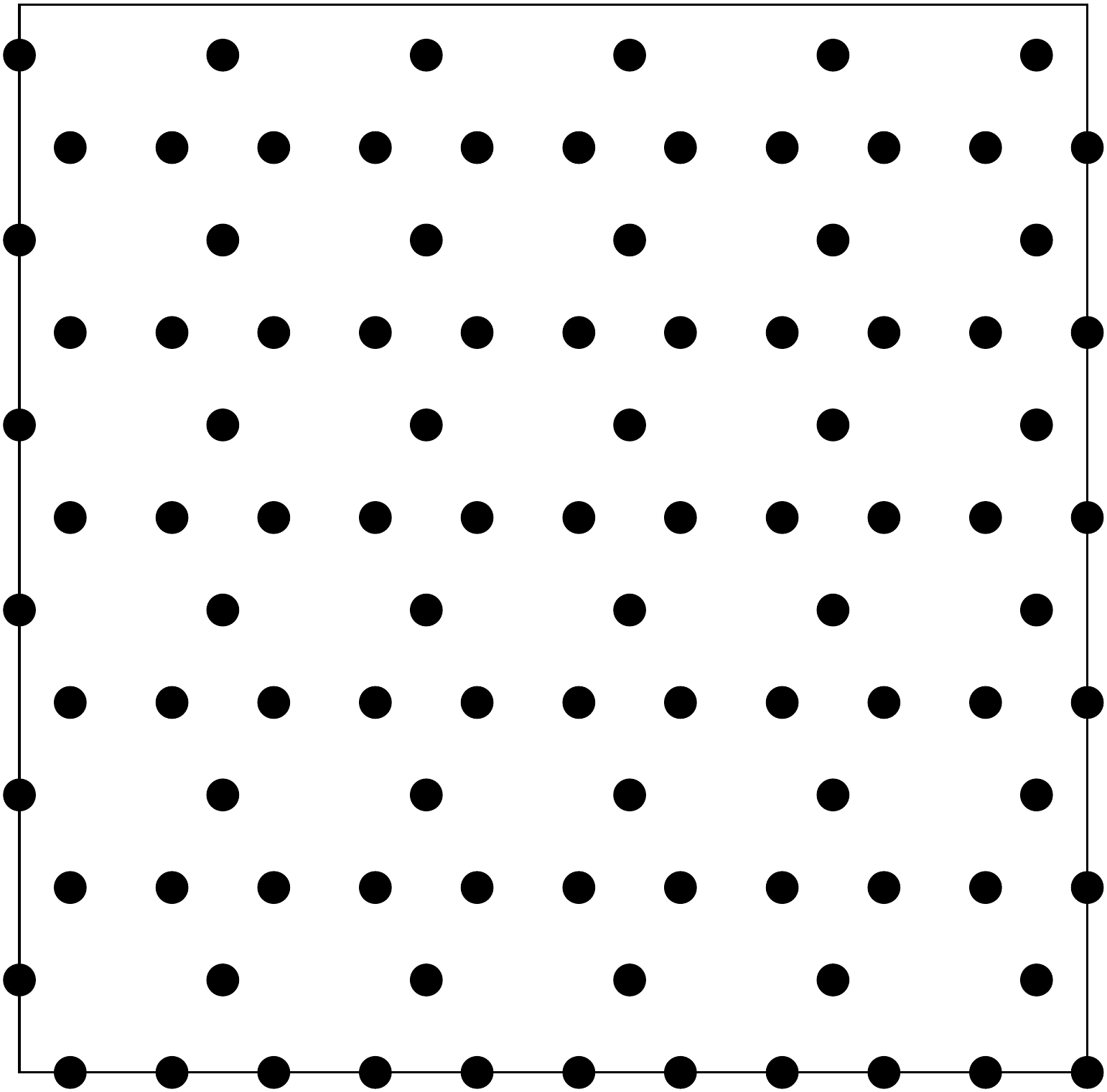}
\caption{\emph{Left:}  The Kagom\' e lattice.  Note that all lattice sites are equivalent up to a rotation.  \emph{Right:}  A four-coordinate
relative of the Kagom\' e lattice with two inequivalent types of lattice sites.}\label{d2latt2}
\end{figure}

\begin{figure}[!tp]
\centering
\includegraphics[width=0.35\textwidth]{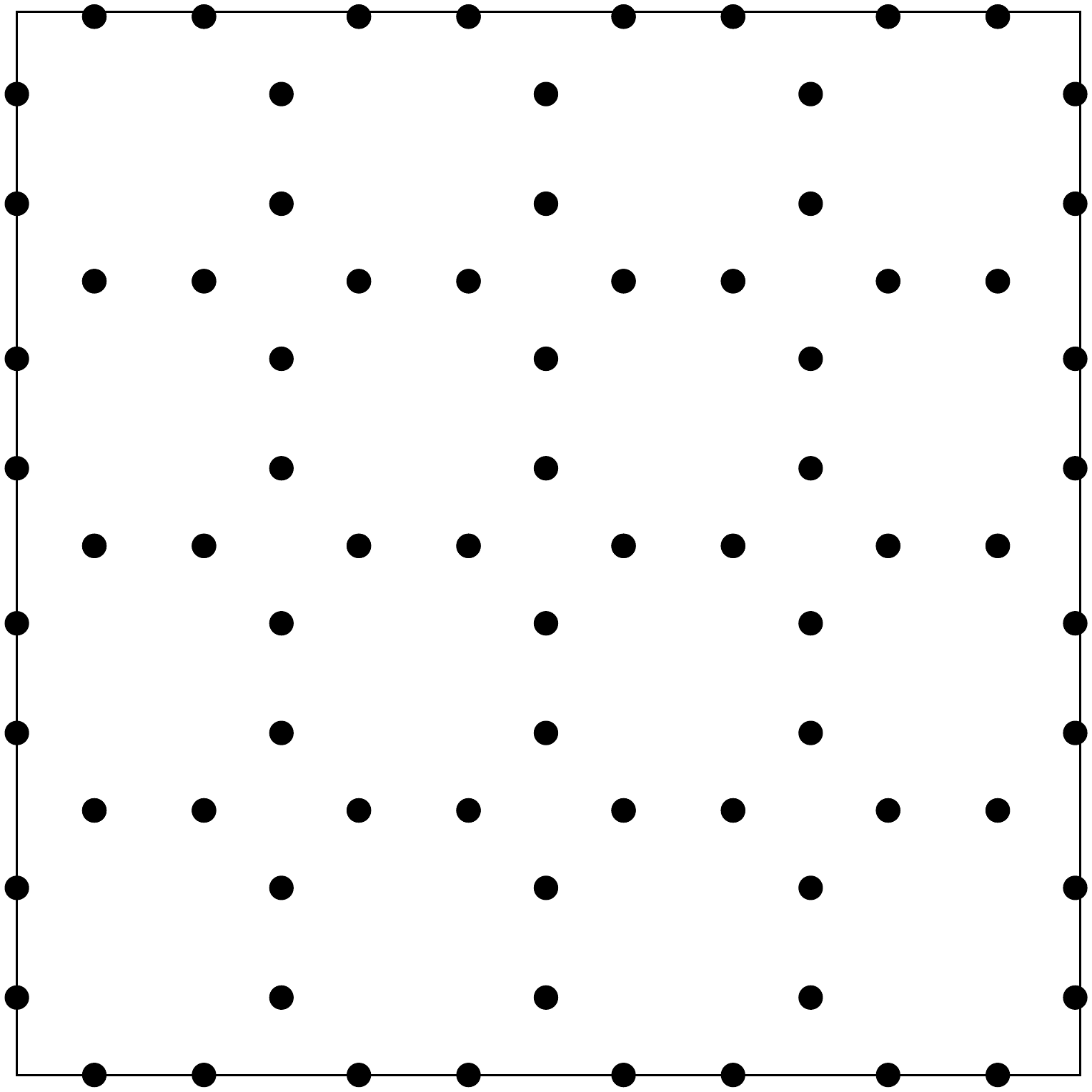}
\caption{A periodic point pattern generated from the $4.8.8$ Archimedean tessellation of $d = 2$ Euclidean space.}\label{archi}
\end{figure}

On account of its long-range order, any periodic point pattern is hyperuniform \cite{ToSt03}.  Although such systems are not strictly statistically homogeneous,
one can still use the theoretical framework developed in the preceding sections by adopting a radially-averaged form of the pair correlation function 
and by ``smoothing'' the long-range correlations to ensure proper convergence of the necessary sums; the details of this procedure are described in Section 5.1. 
Prior work by Torquato and Stillinger \cite{ToSt03} provided some calculations for the asymptotic coefficients $B_N$ of a small number of one-, two-, and 
three-dimensional periodic structures belonging to the $\mathbb{Z}^d, A_d,$ and $D_d$ families of lattices and their duals 
(for a review of common lattices see, e.g., \cite{CoSl99}) in addition to periodic nonlattices and certain disordered systems constructed from so-called ``$g_2$-invariant processes,'' which we describe in more detail below.  
The smallest known values of $B_N$ in these dimensions belong to lattice structures; however, we remark 
that the densest lattices do not necessarily minimize the fluctuations in the number variance.  For example, it is known in three dimensions that the BCC lattice 
possesses a smaller asymptotic coefficient $B_N$ than the FCC lattice \cite{ToSt03}. Indeed, the dual lattices of the densest lattice packings of spheres
are good solutions to the number variance problem \cite{To09}.
In this paper we have extended prior work by Torquato and Stillinger \cite{ToSt03} by calculating asymptotic coefficients $B_N$ for a broader range of 
periodic and quasiperiodic structures.
Figures \ref{d1latt}-\ref{archi} show a few of the lower-dimensional periodic systems that we have studied.  

One set of disordered point processes that can exhibit hyperuniformity is the so-called class of \emph{determinantal point processes}.
A determinantal point process is a stochastic point process with a joint probability distribution given by the determinant of a finite-rank, positive, bounded, and self-adjoint operator.
The determinantal form of the probability distribution is known to 
induce an effective repulsion amongst points in a realization of the corresponding point process, and these systems are therefore ideal candidates for 
hyperuniform point patterns.  Examples of hyperuniform determinantal point processes have been documented both on the real line \cite{CoLe04, CoDuHu08, ToScZa08, ScZaTo09} and in 
higher Euclidean dimensions \cite{ToScZa08, ScZaTo09}.

\begin{figure}[!tp]
\centering
\includegraphics[width=0.4\textwidth]{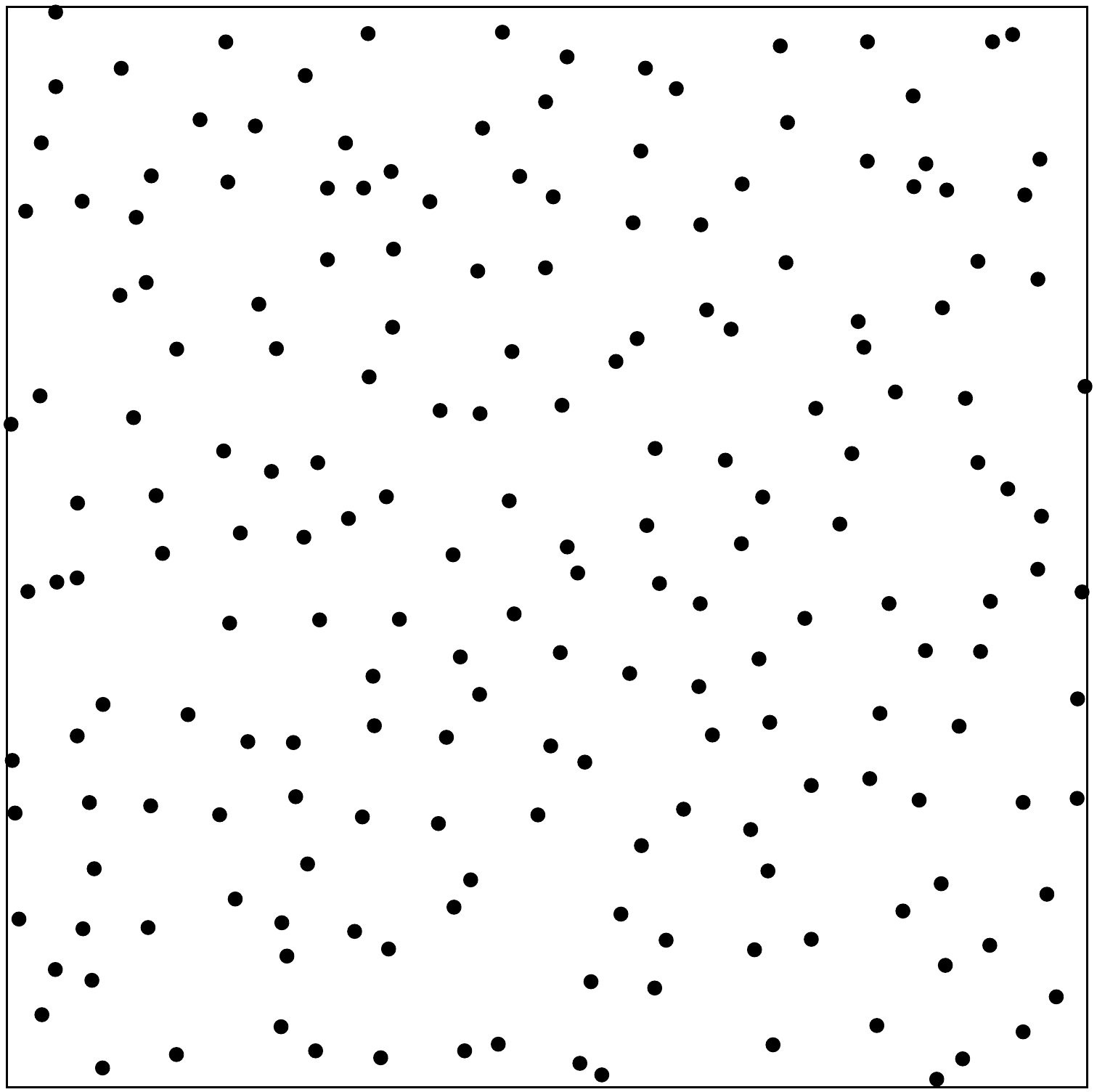}
\caption{The point process generated by the two-dimensional one-component plasma or by the Ginibre ensemble.}\label{ginibrefig}
\end{figure}

A particular example of a determinantal point process that exhibits hyperuniform behavior is generated by the so-called two-dimensional
one-component plasma (see Figure \ref{ginibrefig}); 
the total correlation function for this system
is given by \cite{ToSt03}:
\begin{eqnarray}
h(r) = -\exp\left(-\rho\pi r^2\right),
\end{eqnarray}
and for the case $\rho = 1/\pi$, this system is equivalent to the Ginibre ensemble from random matrix theory \cite{ginibre:440, mehta2004rm, ScZaTo09}.  The corresponding
structure factor has the following small-$k$ behavior \cite{ToSt03}:
\begin{eqnarray}
S(k) \sim k^2 \qquad (k\rightarrow 0),
\end{eqnarray}
and the long-range behavior of the total correlation function $h(r)$ is 
governed by a Gaussian.  This point process extends naturally to higher dimensions but does not retain any connection to random matrix theory.  

Gabrielli and coworkers \cite{GaJaJoLe03} have shown that the so-called ``pinwheel'' tiling \cite{Ra94} of the plane generates a hyperuniform point pattern.  The pinwheel
tiling is constructed by iteratively performing ``decomposition'' and ``inflation'' processes on right triangles with sides of lengths 1, 2, and $\sqrt{5}$.  A point pattern 
is then obtained by randomly placing a point in each elementary triangle, and because the tiles appear in infinitely many orientations, the resulting point process
is statistically homogeneous and isotropic\footnote{There is to date no rigorous proof of homogeneity for the point process generated by the pinwheel tiling; however, 
the current empirical evidence suggests that this claim is likely true \cite{GaJaJoLe03}.  The rotational invariance of the point pattern is experimentally 
manifested by a diffraction pattern consisting of uniform rings rather than isolated Bragg peaks.}.  Also considered by Gabrielli and coworkers is the Harrison-Zeldovich
\cite{Ha70, Ze72} power spectrum for primordial density fluctuations in the Universe; the structure factor here scales linearly for small $k$.  Point patterns in 
three dimensions have been constructed that are consistent with this power spectrum \cite{GaJaJoLe03}.  Gabrielli, Joyce, and Torquato have also provided 
general constructions of hyperuniform point patterns from tilings of $d$-dimensional Euclidean space by appropriately placing $n\geq 1$ points per tile \cite{GaJoTo08}.

Two quasicrystalline structures in the plane that have been studied are the so-called Penrose and octagonal tilings.  The Penrose tiling is composed of obtuse 
and acute rhombi, resulting in an overall five-fold rotational symmetry.  
Similarly, the octagonal tiling is obtained by laying down squares and rhombi in the plane and possesses an overall eight-fold 
rotational symmetry.  These structures have recently received interest in the study and construction of photonic materials \cite{FlToSt09}
that forbid propagation of electromagnetic radiation in some frequency range.  

Torquato, Stillinger, and coworkers have introduced so-called $g_2$-invariant processes, which are ideal methods for constructing hyperuniform point processes from known
realizable pair correlation functions \cite{ToSt02, StToErTr01, SaToSt02}.  A \emph{$g_2$-invariant process} is defined for a chosen nonnegative $g_2$
via analytic extension with invariant form over a nonvanishing density range while keeping all other relevant macroscopic variables fixed.  The upper limiting 
``terminal'' density is the point above which the nonnegativity condition on the structure factor is violated, implying that at the terminal density the system 
is hyperuniform if realizable.  Torquato and Stillinger \cite{ToSt03} have calculated the asymptotic coefficients for a number of $g_2$-invariant processes,
and for completeness we include their results below in Tables \ref{d1table}, \ref{tableone}, and \ref{tabletwo}.

We finally remark that is has been conjectured that all saturated strictly jammed packings of hard spheres generate hyperuniform point patterns \cite{ToSt03, DoStTo05}.   
A \emph{strictly jammed} sphere packing is rigorously incompressible and nonshearable, and a \emph{saturated} packing is one for which there is no space available to 
add another sphere.  Of particular interest in this regard is the so-called \emph{maximally random jammed} (MRJ) state, defined as the most disordered system
(according to some well-defined order metric) among all strictly jammed packings \cite{ToTrDe00}.  Donev and coworkers \cite{DoStTo05} have provided strong 
numerical evidence to suggest that this disordered system is hyperuniform at saturation with a structure factor $S(k)$ that is linear in $k$ as $k\rightarrow 0$, suggesting that the conjecture is likely true. 

\subsection{New hyperuniform point patterns}

We mention here the so-called ``tunneled'' crystals introduced by Torquato and Stillinger \cite{ToSt07}; these periodic structures are formed by stacking two-dimensional
honeycomb layers according to some repeated pattern in a manner similar to the construction of the FCC and HCP structures from triangular layers.  The resulting crystals 
are unsaturated and yet hyperuniform due to the periodicity enforced by the stacking order \cite{ToSt07}.  The tunneled FCC crystal contains a periodic array of parallel 
linear tunnels that can be oriented in any one of six equivalent directions; in contrast, the tunneled HCP crystal contains tunnels with an overall zig-zag shape 
with three possible lateral directional orientations.  As with the saturated parent packings, the tunneled FCC and tunneled HCP crystals have the same packing fractions;
for more details on these systems, see \cite{ToSt07}.  

Another novel application of this work in one dimension 
has involved the characterization of the so-called \emph{Fibonacci chain}, which is a prototypical quasicrystal.  A \emph{quasicrystal} is a nonperiodic 
system whose correlation structure (i.e., $g_2$ and $S$) exhibits Bragg-like behavior, implying the presence of long-range order 
despite the absence of translational symmetry (see \cite{LeSt84} and \cite{Ja94}).  The Fibonacci chain is specifically composed of two length segments (``long'' $L$ and 
``short'' $S$) distributed iteratively on the line according to the following substitution rules:  $S\rightarrow L$ and $L \rightarrow LS$.  Upon taking the 
thermodynamic limit one obtains a quasiperiodic tiling of space; a point process can be generated by taking the endpoints of each segment (see Figure \ref{d1latt}).

\begin{figure}[!tp]
\centering
\includegraphics[width=0.25\textwidth]{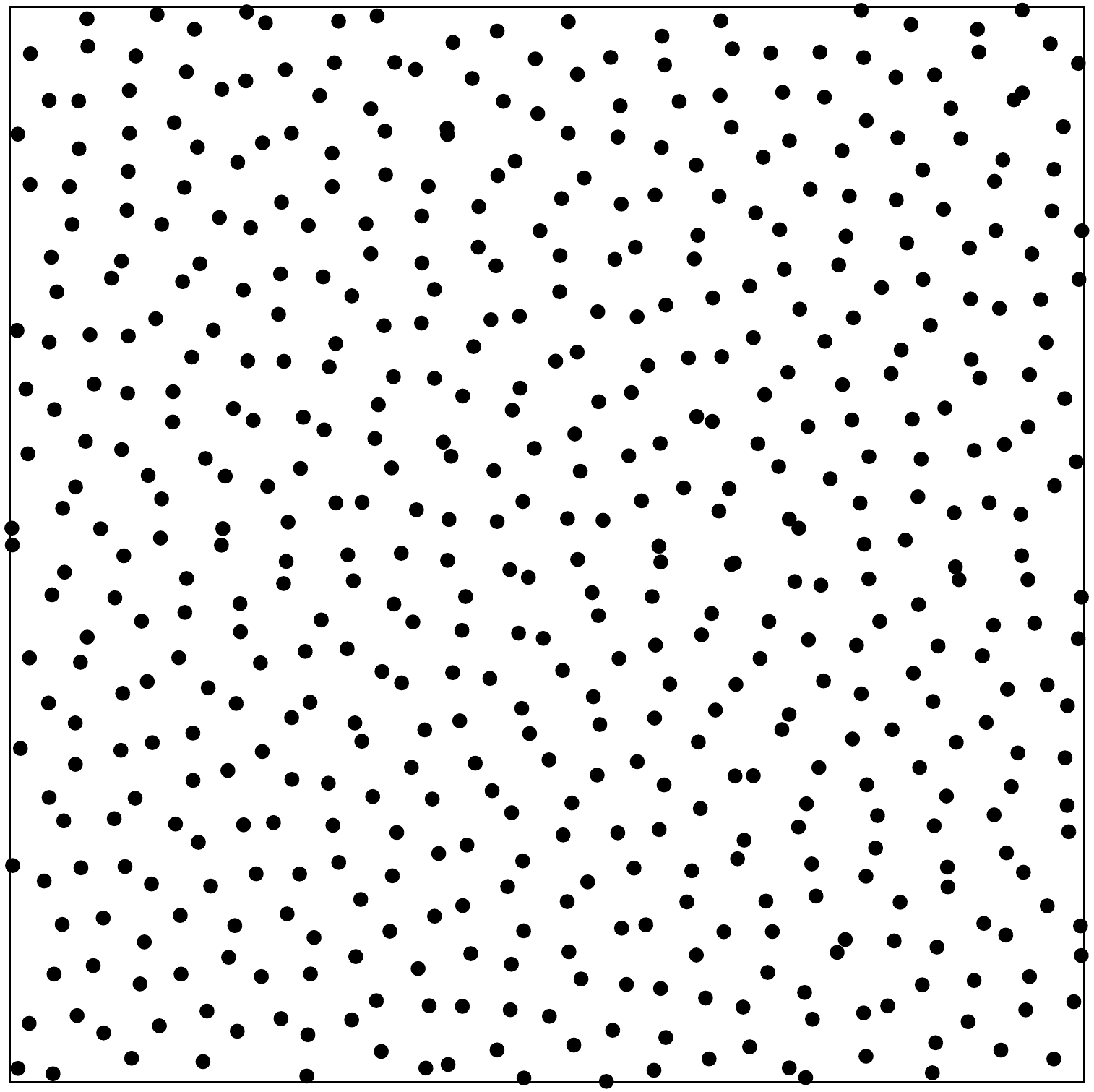}\hspace{0.05\textwidth}
\includegraphics[width=0.25\textwidth]{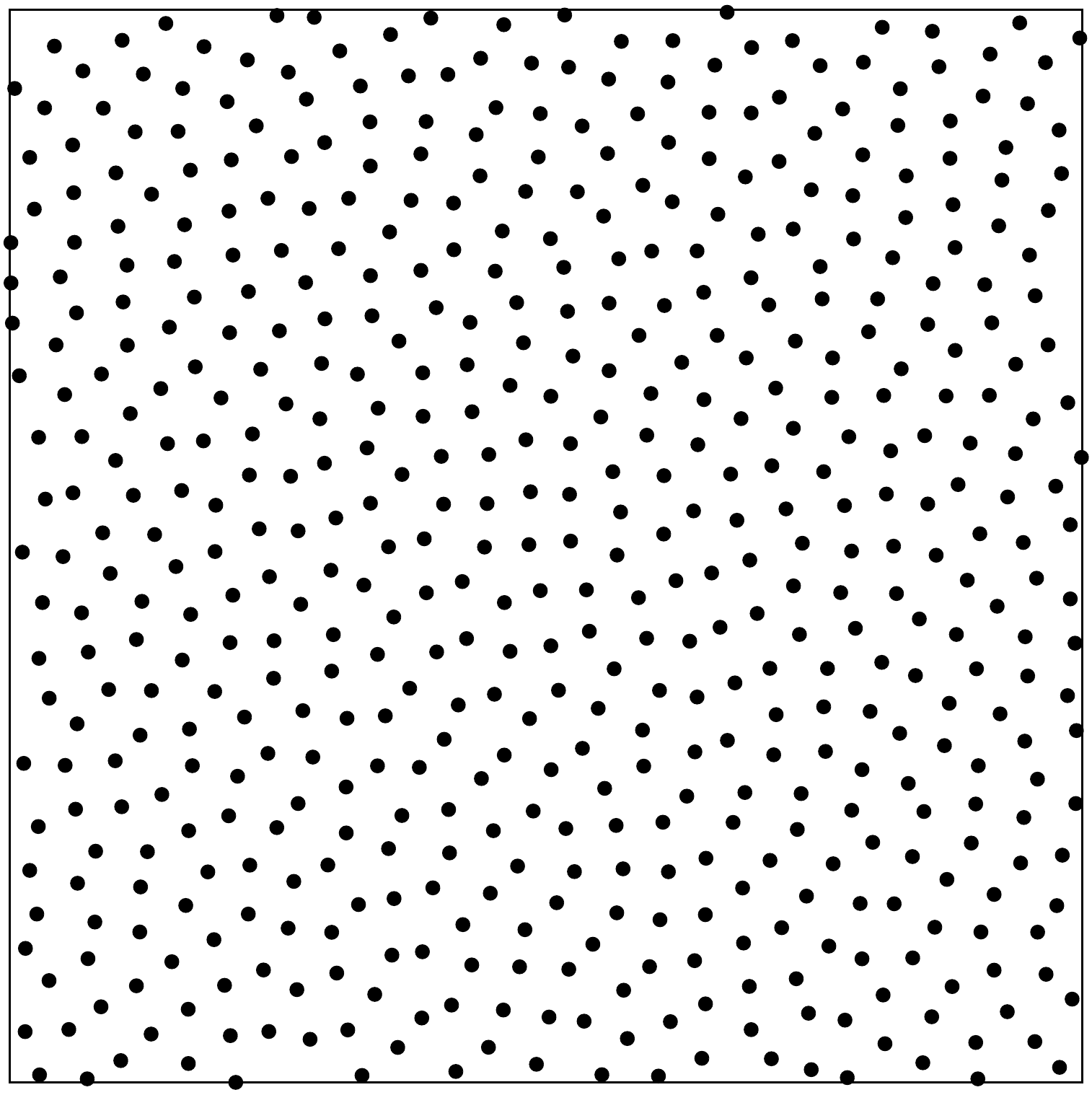}\hspace{0.05\textwidth}
\includegraphics[width=0.25\textwidth]{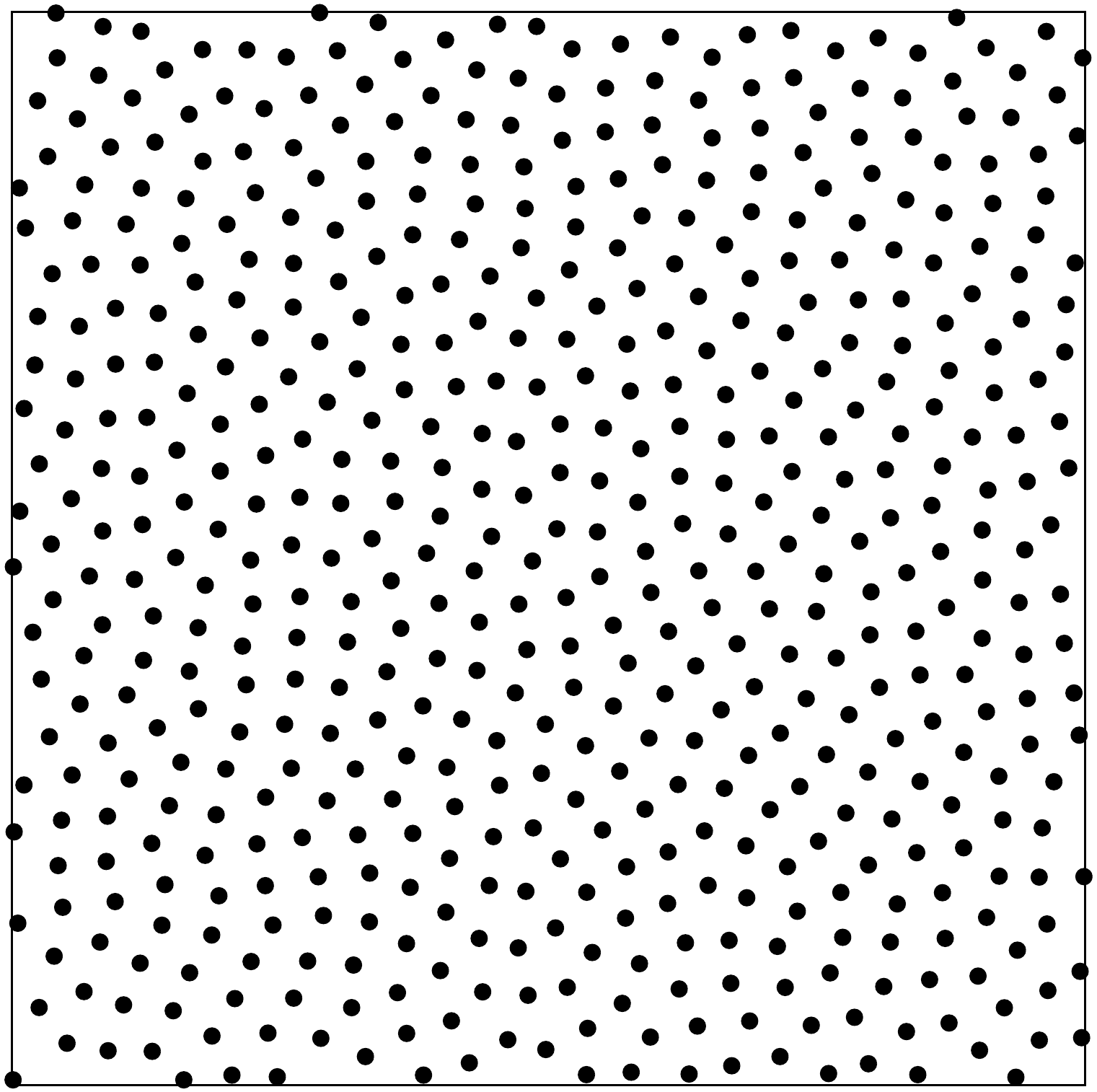}
\caption{Hyperuniform classical disordered ground states obtained by minimizing the collective coordinate
representation of the pairwise interaction $\Phi$ in (\ref{collcoord}).  The fraction of constrained degrees of freedom $\gamma$ is given by: (\emph{left}) $\gamma = 0.302$;
(\emph{center}) $\gamma = 0.402$; (\emph{right}) $\gamma = 0.496$.}\label{groundstates}
\end{figure}

With regard to disordered hyperuniform point patterns, 
Batten, Stillinger, and Torquato  \cite{BaStTo08} have numerically constructed examples of hyperuniform classical disordered ground states; these systems,
which they describe as ``stealth'' materials, 
are prototypical 
heterogeneous media that completely suppress scattering of incident radiation for a specified set of wavevectors.  The construction of these systems 
relies upon the minimization of a pairwise interaction $\Phi$, which can be represented in Fourier space as \cite{BaStTo08}:
\begin{eqnarray}\label{collcoord}
\Phi = \vert\Omega\vert^{-1} \sum_{\mathbf{k}} \hat{v}(\mathbf{k}) C(\mathbf{k}),
\end{eqnarray}
where $\vert\Omega\vert$ denotes the Lebesgue measure of the system domain, $\hat{v}(\mathbf{k})$ is the Fourier transform of a 
translationally invariant two-particle potential energy $v(\mathbf{x}-\mathbf{y})$, and $C(\mathbf{k})$ is defined by:
\begin{eqnarray}\label{Ck}
C(\mathbf{k}) = \frac{N}{2}\left[S(\mathbf{k})-1\right].
\end{eqnarray}
The function $\hat{v}(\mathbf{k})$ is chosen to be positive and bounded with compact support on some interval $\Vert\mathbf{k}\Vert \in [0, K]$; the authors 
in \cite{BaStTo08} also
choose this function as isotropic, but the specific form of $\hat{v}(\mathbf{k})$ is largely inconsequential so long as it satisfies the properties enumerated above.  
By construction, the minimum of $\Phi$ is then determined by minimizing $C(\mathbf{k})$ in the interval $[0, K]$, which by (\ref{Ck}) requires that $S(\mathbf{k}) = 0$
for all such wavenumbers.  Therefore, it is clear that this construction algorithm inherently generates a hyperuniform point pattern.  We remark that the 
degree of order in the resulting system is dependent on the ratio $\gamma$ of the number of constrained degrees of freedom to the total number of degrees of freedom\footnote{The authors in \cite{BaStTo08} refer 
to the ratio that we denote here as $\gamma$ by the letter $\chi$.  Since we have reserved the latter notation 
for the autocovariance function associated with the two-point probability $S_2$, we adopt a different convention in this paper.}.  Figure \ref{groundstates}
provides sample realizations of these hyperuniform disordered ground states for several values of $\gamma$; note that the degree of global order in the system
increases with increasing $\gamma$.

\section{Higher-dimensional number variance coefficients and local volume fraction estimates}

Here we provide extensive tables of 
asymptotic number variance coefficients for several Bravais lattice families up to $d = 8$, and these results, where appropriate, are compared to 
certain nonlattice ground states introduced by Cohn, Kumar, and Sch\"urmann \cite{CoKu09}.  Estimates for the asymptotic volume fraction
variance coefficients of the lattice systems may therefore be developed, 
providing insight into the dimensional asymptotic behavior of $B_{\tau}$.  

High-dimensional systems play an integral role in statistical physics.  Of particular interest in this regard is the so-called sphere-packing problem, which 
has applications to abstract algebra, number theory, and communications theory \cite{CoSl99}.  For example, it is known that the optimal method of sending digital signals over noisy
channels corresponds to the densest sphere packing in a
high-dimensional space.  Torquato
and Stillinger have also provided evidence for a decorrelation
principle of disordered packings in high-dimensional Euclidean spaces \cite{ToSt06}. This principle states that as the dimension $d$
increases, all unconstrained correlations vanish, and any
higher-order correlation functions $g_n(\mathbf{x}_1 , \ldots ,\mathbf{x}_n)$ may be
written in terms of the number density $\rho$ and the pair correlation function
function $g_2$ within some small error.  The decorrelation principle appears to be robust, apparently incorporating
a large number of point processes beyond sphere packings \cite{ZaStTo08, ToScZa08, ScZaTo09}.  As a result, it is likely that glassy states of matter
are intrinsically more stable than crystals in high dimensions and pack with a greater density \cite{ToSt06}, which should be contrasted 
with physical systems in two- and three-dimensions.

It is instructive to compare the estimate (\ref{BNtau}) for the asymptotic coefficient $B_{\tau}$ to an exact result for an analytically soluble system.  We therefore
consider the so-called ``step-function'' point process, defined at the two-point level (and therefore not uniquely) by:
\begin{eqnarray}\label{stepfunction}
g_2(r) = \Theta(r-D),
\end{eqnarray}
where $\Theta$ is the Heaviside step function and $D$ is an effective diameter for the point process.  Note that the form of the pair correlation function (\ref{stepfunction})
is exactly realized for an equilibrium system of hard spheres with diameter $D$ in the limit of low density.  However, there is also strong theoretical and 
numerical evidence \cite{SaToSt02, StToErTr01, CrToSt03} to suggest that this pair correlation is realizable for nonzero densities up to the critical value 
$\phi_c = 1/2^d$.  It is at this critical density that the structure factor approaches zero for small $k$, and the system becomes hyperuniform\footnote{Above the 
critical density the structure factor for the step-function process falls below zero, signifying a violation of the Wiener-Khinchtine condition.  The pair correlation
function then ceases to be realizable as a point process \cite{ToSt03}.}.  
The corresponding total correlation function $h$ is given by:
\begin{eqnarray}\label{steph}
h(r) = -\Theta(D-r),
\end{eqnarray}
which admits the Fourier representation:
\begin{eqnarray}\label{stephfourier}
\hat{h}(k) = -\left(\frac{2\pi d}{k}\right)^{d/2} J_{d/2}(Dk).
\end{eqnarray}

The asymptotic coefficient for the number variance $B_N$ at the critical density can be calculated analytically for this process; taking the $d$-th moment of the total
correlation function (\ref{steph}) and using the definition (\ref{Bn}), we find:
\begin{eqnarray}\label{stepBn}
B_N = \frac{d\Gamma(1+d/2)}{\sqrt{\pi}(d+1) \Gamma\left[(d+1)/2\right]}.
\end{eqnarray}
The dimensional asymptotic behavior of this coefficient follows directly from the corresponding behavior of the gamma function; namely:
\begin{eqnarray}\label{Bnstepscale}
B_N \sim \sqrt{\frac{d}{2\pi}} + \mathcal{O}(1/\sqrt{d}) \qquad (d\rightarrow +\infty).
\end{eqnarray}
It is interesting to note that the pair correlation function (\ref{stepfunction}) is also realizable as the infinite-dimensional limit of the so-called 
``ghost'' random sequential addition (RSA) process at the saturation density $\phi_c = 1/2^d$ \cite{ToSt06c}.  This model is a generalization of the well-studied
RSA process in which points are placed sequentially in some space according to a uniform distribution subject to the constraint that the points do not fall within 
some prescribed distance of either other points in the system \emph{or} ``ghost'' points representing prior failed particle insertions.  We remark that the resulting 
process represents a \emph{nonequilibrium} hard-sphere packing.  The result in (\ref{Bnstepscale}) suggests a high-dimensional divergence in the local number
variance for this system, and we more thoroughly explore the implications of this behavior at the end of this section.

An explicit calculation of the asymptotic coefficient $B_{\tau}$ can also be done for the step-function point process in low dimensions.  The most difficult aspect of this calculation 
is the determination of the convolution $(h*v_{\mbox{int}})(r; \lambda)$ in (\ref{autocorrelHS}); 
an Fourier representation of this function is obtained by taking 
the Fourier transform of the convolution using (\ref{stephfourier}) and the result:
\begin{eqnarray}
\hat{v}_{\mbox{int}}(k; \lambda) = \left(\frac{2\pi\lambda}{k}\right)^{d} \left[J_{d/2}(k\lambda)\right]^2.
\end{eqnarray}
We therefore find:
\begin{eqnarray}\label{stepchi}
\chi(r) &= \rho v_{\mbox{int}}(r; \lambda)-\rho^2 \int_0^{+\infty} k^{d-1} \left(\frac{D}{k}\right)^{d/2} \left(\frac{2\pi\lambda}{k}\right)^d\times\nonumber\\
&\left\{\frac{J_{d/2}(Dk)\left[J_{d/2}(k\lambda)\right]^2 J_{d/2-1}(kr)}{(kr)^{d/2-1}}\right\} dk.
\end{eqnarray}
We remark that there is no reason \emph{a priori} to identify the particle radius $\lambda$ with $D/2$; the length scale $D$ is imposed inherently by the structure
of the underlying point process and is not affected by a ``decoration'' of the points via circumscription with spheres.  However, it is computationally convenient 
to make this connection, and we additionally set $D = 2\lambda = 1$. 

For $d = 1$, the expression (\ref{stepchi}) simplies according to:
\begin{eqnarray}\label{stepchi2}
\fl \chi(r) = \phi\left(1-r\right)\Theta(1-r)+\frac{\rho^2}{4}\left[r^2-2+(r-2)\vert r-2\vert-2(r-1)\vert r-1\vert\right],
\end{eqnarray}
where $\rho = \phi = 1/2$ at the critical density.
Using (\ref{btau}) we find the following result for the asymptotic coefficient:
\begin{eqnarray}
B_{\tau} = 1/16.
\end{eqnarray}
One can also verify directly from (\ref{stepchi2}) that $A_{\tau} = 0$ as expected for a hyperuniform medium.  According to (\ref{Bnstepscale}), the corresponding
coefficient for the local number density is,
\begin{eqnarray}
B_N = 1/4;
\end{eqnarray}
and applying the estimate (\ref{BNtau}) leads to the following upper bound on $B_{\tau}$:
\begin{eqnarray}
B_{\tau} \leq 1/8,
\end{eqnarray}
which we see is strictly fulfilled for this system.  

\begin{figure}[!htp]
\centering
\includegraphics[width=0.40\textwidth]{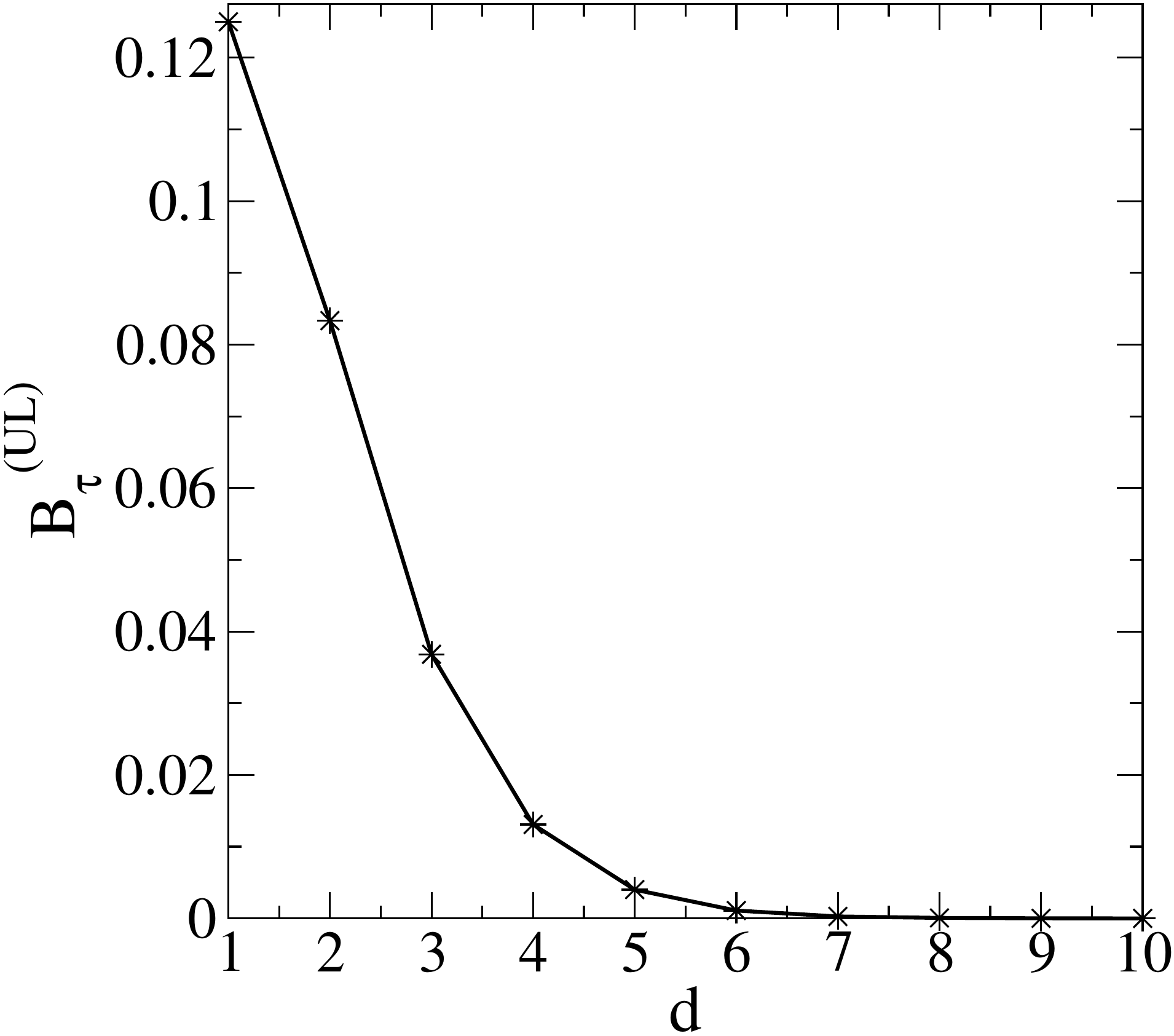}
\caption{Estimate (\ref{BNtau}) for the asymptotic coefficient $B_{\tau}$ of the step-function point process for increasing dimensions.}\label{Btest}
\end{figure}

It is therefore clear that the upper limit in (\ref{BNtau}) will generally not be obtained for a given random medium; however, this estimate does provide
some insight into the dimensional asymptotic properties of $B_{\tau}$.  Figure \ref{Btest} shows the behavior of the estimate (\ref{BNtau}) for the step-function 
process
as the dimension of the system increases.  We observe that, unlike the divergent behavior in the coefficient $B_N$, $B_{\tau}$ decreases at most monotonically
as the dimension increases and vanishes asymptotically for large dimensions\footnote{It should be noted that monotonicity of the upper bound
(\ref{BNtau}) for $B_{\tau}$ is not generally true for all hyperuniform systems.  However, this issue does not affect our conclusions concerning the 
dimensional asymptotic behavior of this estimate.}. 
It is known that fluctuations in the local number density in a 
spherical observation window cannot grow more 
slowly than the surface area of the window \cite{Be87, ToSt03}, and if this claim is also true for local volume fraction fluctuations, then it follows 
that in high dimensions the local volume fraction for large windows is almost surely equal to the global volume fraction in the system. 


\begin{figure}[!htp]
\centering
\includegraphics[width=0.40\textwidth]{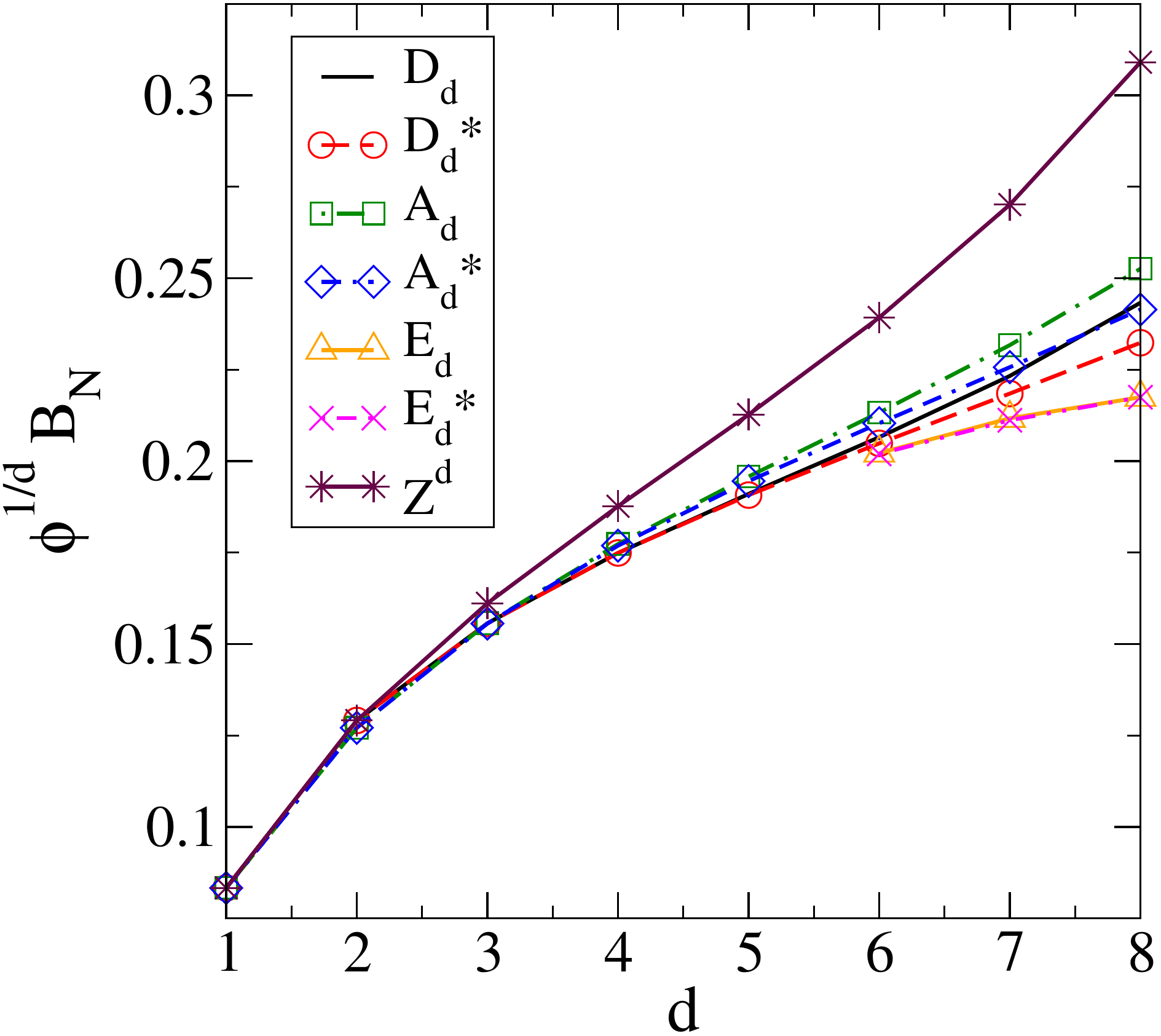}\hspace{0.05\textwidth}
\includegraphics[width=0.40\textwidth]{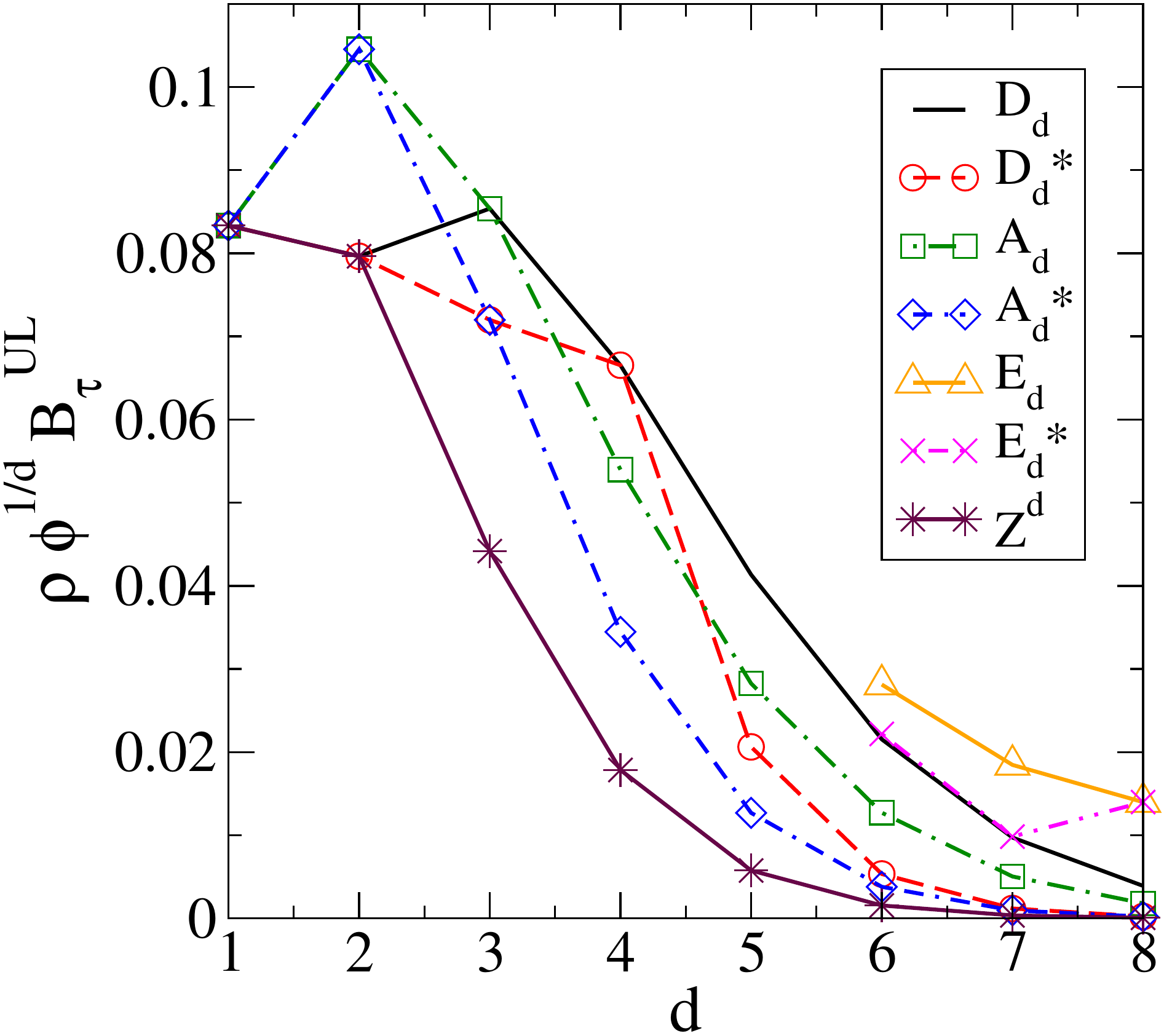}
\caption{\emph{Left:} Dimensional scaling of the asymptotic number variance coefficient $B_N$ for several families of lattices.\\
\emph{Right:}  Corresponding scaling for the estimate of the asymptotic local volume fraction variance coefficient $B_{\tau}$.}\label{lattfig}
\end{figure}

One expects that this 
behavior is quite general for other disordered systems by invoking a so-called \emph{decorrelation principle}.  
As discussed by Torquato and Stillinger \cite{ToSt06},
the decorrelation principle states that in the high-dimensional limit, \emph{unconstrained} correlations asymptotically vanish, and knowledge about 
higher-order correlations ($g_n$ for $n\geq 3$) can be inferred directly from knowledge of the pair correlation function $g_2$ and the number density $\rho$. 
In other words, long-range fluctuations in $g_2(r)$, and therefore $\chi(r)$, should diminish as the dimension of the system increases; since
these fluctuations are responsible for enforcing the constraint:
\begin{eqnarray}
\int_0^{+\infty} r^d \chi(r) dr < 0,
\end{eqnarray}
it follows that the value of this integral becomes sequentially \emph{less negative} in higher dimensions.  Therefore, we expect that for general 
heterogeneous media, including but not limited to the step-function process above, $B_{\tau} \rightarrow 0$ as $d \rightarrow +\infty$ in constrast 
to the behavior of $B_N$.  It is in this sense that we claim that high-dimensional hyperuniform heterogeneous media can be generated from 
increasingly disordered point processes and yet possess inherently ordered microstructural information in terms of the local volume fraction.

\Table{\label{latt-table}  Asymptotic scaling coefficients $\phi^{1/d} B_N$ for fluctuations in the local number density of several lattice families.  Also included 
are the maximal packing fractions $\phi$ of the lattices in each dimension.  Note that the $E_d$ and $E_d^*$ lattice families are only uniquely defined
for $d\geq 6$; a hyphen therefore indicates that information for these lattices is not available in lower dimensions.}
\br
$d$ & $\mathbb{Z}^d$ $(\phi)$ & $A_d$ $(\phi)$ & $A_d^*$ $(\phi)$\\
\mr
1 & 0.08333 (1) & 0.08333 (1) & 0.08333 (1)\\
2 & 0.12910 ($\pi/4$) & 0.12709 ($\sqrt{3}\pi/6$) & 0.12709 ($\sqrt{3}\pi/6$)\\
3 & 0.16115 ($\pi/6$) & 0.15569 ($\sqrt{2}\pi/6$) & 0.15560 ($\sqrt{3}\pi/8$)\\
4 & 0.18767 ($\pi^2/32$) & 0.17733 ($\sqrt{5}\pi^2/40$) & 0.17689 ($\sqrt{5}\pi^2/50$)\\
5 & 0.21273 ($\pi^2/60$) & 0.19579 ($\sqrt{3}\pi^2/45$) & 0.19453 ($5\sqrt{5}\pi^2/432$)\\
6 & 0.23925 ($\pi^3/384$) & 0.21327 ($\sqrt{7}\pi^3/336$) & 0.21036 ($9\sqrt{7}\pi^3/5488$)\\
7 & 0.27012 ($\pi^3/840$) & 0.23170 ($\pi^3/210$) & 0.22569 ($49\sqrt{7}\pi^3/61440$)\\
8 & 0.30904 ($\pi^4/6144$) & 0.25258 ($\pi^4/1152$) & 0.24144 ($2\pi^4/6561$)\\
\br
$d$ & $D_d$ $(\phi)$ & $D_d^*$ $(\phi)$ & $E_d$ $(\phi)$\\
\mr
1 & 0.08333 (1) & 0.08333 (1) & - \\
2 & 0.12910 ($\pi/4$) & 0.12910 ($\pi/4$) & -  \\
3 & 0.15569 ($\sqrt{2}\pi/6$) & 0.15560 ($\sqrt{3}\pi/8$) & - \\
4 & 0.17488 ($\pi^2/16$) & 0.17488 ($\pi^2/16$) & - \\
5 & 0.19103 ($\sqrt{2}\pi^2/30$) & 0.19070 ($\pi^2/30$) & - \\
6 & 0.20653 ($\pi^3/96$) & 0.20485 ($\pi^3/192$) & 0.20221 ($\sqrt{3}\pi^3/144$) \\
7 & 0.22327 ($\sqrt{2}\pi^3/210$) & 0.21845 ($\pi^3/420$) & 0.21172 ($\pi^3/105$) \\
8 & 0.24329 ($\pi^4/768$) & 0.23236 ($\pi^4/3072$) & 0.21746 ($\pi^4/384$) \\
\br
$d$ & $E_d^*$ $(\phi)$ &  & \\
\mr
1 & -  &  & \\
2 & - &  & \\
3 & -  &  & \\
4 & - &  & \\
5 & - &  & \\
6 & 0.20195 ($\sqrt{3}\pi^3/162$) & & \\
7 & 0.21119 ($9\sqrt{3}\pi^3/2240$) & & \\
8 & 0.21746 ($\pi^4/384$) & &\\
\br
\endTable

The analysis of periodic structures is seemingly more complicated.  These systems inherently contain long-range order, and there is no 
reason \emph{a priori} for a decorrelation principle to apply. 
Nevertheless, the same qualitative trends are observed for these systems as seen in Figure \ref{lattfig}, which 
compares the asymptotic coefficients $B_N$ and $B_{\tau}$ for 
several lattice families across dimensions; the results for the number variance coefficients are also collected in Table \ref{latt-table}.  
Note that for a meaningful comparison among lattice families we have scaled the coefficients appropriately 
to remove the dependence on the parameter $D$, which amounts to mapping $B_N \rightarrow \phi^{1/d} B_N$ and $B_{\tau} \rightarrow \rho \phi^{1/d} B_{\tau}$
according to (\ref{numasymp}) and (\ref{tauasymp}); the factor of $\rho$ is included in the rescaling of $B_{\tau}$ for computational convenience.  
As with the step-function process, the coefficients for the local number density increase monotonically and 
likely diverge asymptotically with respect to increasing dimension.
Fluctuations in the local volume fraction generally decrease for sufficiently large dimensions, although not necessarily monotonically.  
It is interesting to note that the asymptotic number variance coefficient for a Bravais lattice is apparently larger than its dual, 
and this trend also appears to hold for the local volume fraction.
However, it seems that the integer lattices $\mathbb{Z}^d$ may surprisingly possess the smallest volume fraction fluctuations among these lattices despite
having significantly smaller packing fractions.  This behavior arises mathematically from the rapidly diminishing volume fraction for the integer lattice in
high dimensions, which has the effect of drastically lowering the upper bound to the scaled local volume fraction asymptotic coefficient relative to other lattices.
We remark, however, that the curves in Figure \ref{lattfig} are simply estimates of the volume fraction fluctuations, meaning that global minimization of $B_{\tau}$ 
cannot be readily ascertained from these results.  
However, we can conclude that it is not necessarily true 
that minimizing fluctuations in the local number density incidentally minimizes fluctuations in the local volume fraction for a 
heterogeneous medium. 

Cohn and Kumar \cite{CoKu08} have also discussed nonlattice packings that possess lower ground-state energies than the densest Bravais lattices
for certain classes of soft-core pair potentials.  
Specifically, they consider in five and seven dimensions certain stacking variants, first introduced by Conway and Sloane \cite{CoSl95},
of the densest Bravais lattices that possess lower ground state energies for the so-called Gaussian core model, which involves a pair potential 
$v(r) = \epsilon\exp\left[-(r/\sigma)^2\right]$.  We adopt their notation for these energy-minimizing nonlattices $\Lambda_5^2$ $(d = 5)$ and $\Lambda_7^3$ $(d = 7)$.  Since the Gaussian core potential and the overlap potential governing the number variance are both completely monotonic (the former in the squared distance $r^2$), we expect the ground-state structures of these interactions to be similar \cite{ToSt08}.  With this notion in mind we mention that the nonlattice packings $\Lambda_5^2$ and $\Lambda_7^3$ actually belong to one-parameter families $D_5^+(\alpha)$ and $D_7^+(\alpha)$ \cite{CoKu09} of packings, the theta series for which are given by:
\begin{eqnarray}
\Theta_{D_5^+(\alpha)}(q) &= \frac{1}{2}\left[\left(\theta_3(q)\right)^4 + \left(\theta_4(q)\right)^4\right]\theta_3(q^{4\alpha^2})\nonumber\\
&+ \frac{1}{6}\left[\left(\theta_3(\sqrt{q})\right)^4 + \left(\theta_4(\sqrt{q})\right)^4-\left(\theta_3(q)\right)^4-\left(\theta_4(q)\right)^4\right]\times\nonumber\\
&\left[\theta_3(q^{\alpha^2/4}) - \theta_3(q^{4\alpha^2})\right]\\
\Theta_{D_7^+(\alpha)}(q) &= \frac{1}{2}\left[\left(\theta_3(q)\right)^6 + \left(\theta_4(q)\right)^6\right]\theta_3(q^{4\alpha^2})\nonumber\\ 
&+ \frac{1}{2}\left[\left(\theta_3(q)\right)^6-\left(\theta_4(q)\right)^6\right]\left[\theta_3(q^{\alpha^2})-\theta_3(q^{4\alpha^2})\right]\nonumber\\
&+\frac{1}{2}\left(\theta_2(q)\right)^6\left[\theta_3(q^{\alpha^2/4})-\theta_3(q^{\alpha^2})\right].
\end{eqnarray}
The nonlattices $\Lambda_5^2$ and $\Lambda_7^3$ are obtained by setting $\alpha = 2$ and $\alpha = \sqrt{2}$, respectively.  
Following Cohn, Kumar, and Sch\"urmann \cite{CoKu09}, we define two structures to be \emph{formally dual} if their pair correlation functions have the same relationship as Poisson summation gives 
for a Bravais lattice and its reciprocal lattice.
For the families $D_5^+(\alpha)$ and $D_7^+(\alpha)$, formal duality is obtained by replacing $\alpha$ with $1/\alpha$; this statement implies that formally self-dual packings occur for $\alpha = 1$.  

\Table{\label{nonlatt_table}  Asymptotic scaling coefficients $\phi^{1/d} B_N$ for fluctuations in the local number density of  nonlattices from the $D_5^+(\alpha)$ $(d = 5)$ and $D_7^+(\alpha)$ $(d= 7)$ one-parameter families.}
\br
Packing & $\phi^{1/d} B_N$\\
\mr
$\Lambda_5^2$ $(\alpha = 2)$ & 0.19099\\
$\Lambda_5^{2*}$ $(\alpha = 1/2)$ & 0.19069\\
$D_5^+(1)$ & 0.21037\\
$\Lambda_7^3$ $(\alpha = \sqrt{2})$ & 0.21115\\
$\Lambda_7^{3*}$ $(\alpha = 1/\sqrt{2})$ & 0.21090\\
$D_7^+(1)$ & 0.21037\\
\br
\endTable

\begin{figure}[!htp]
\centering
\includegraphics[width=0.45\textwidth]{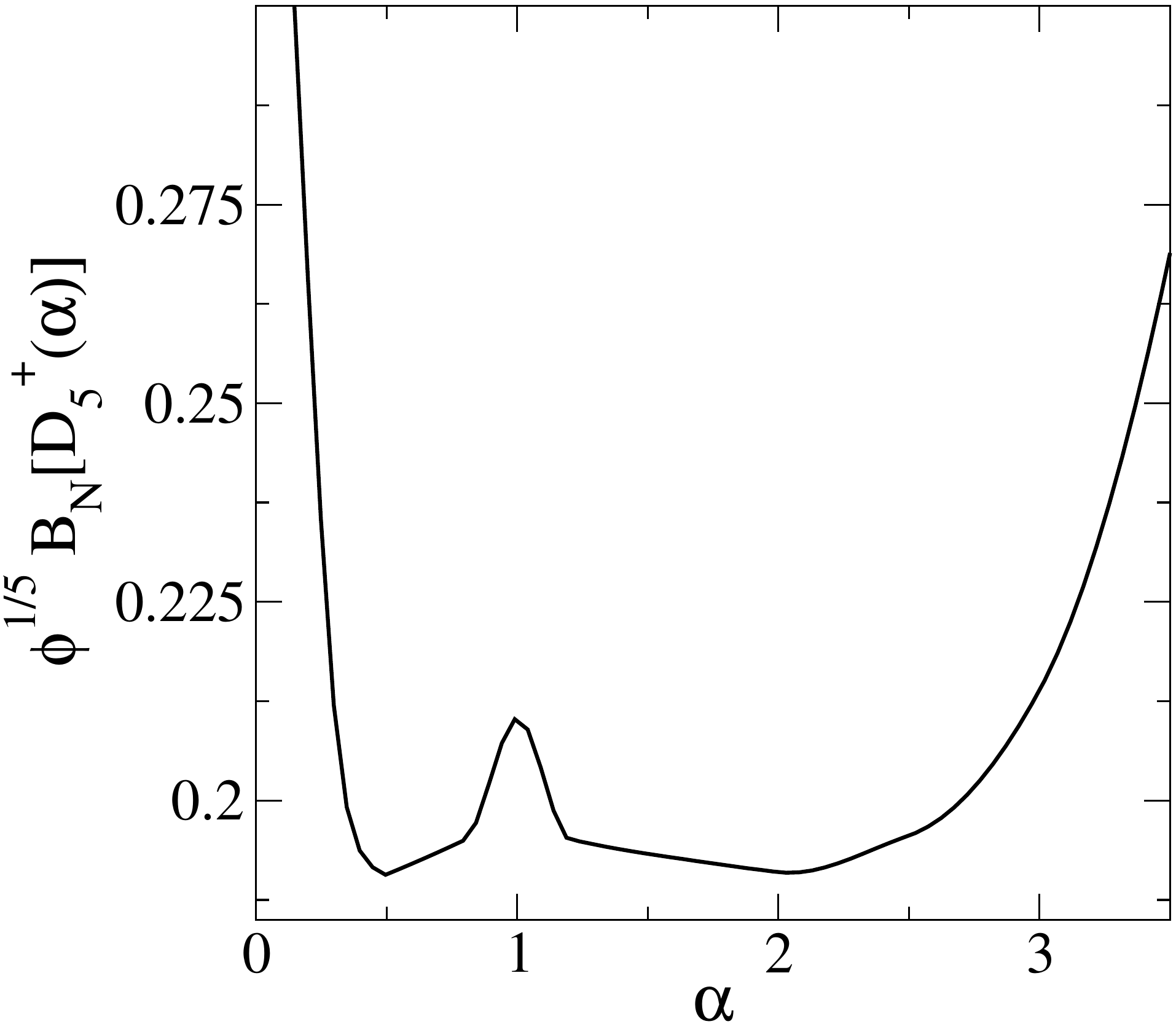}\hspace{0.05\textwidth}
\includegraphics[width=0.45\textwidth]{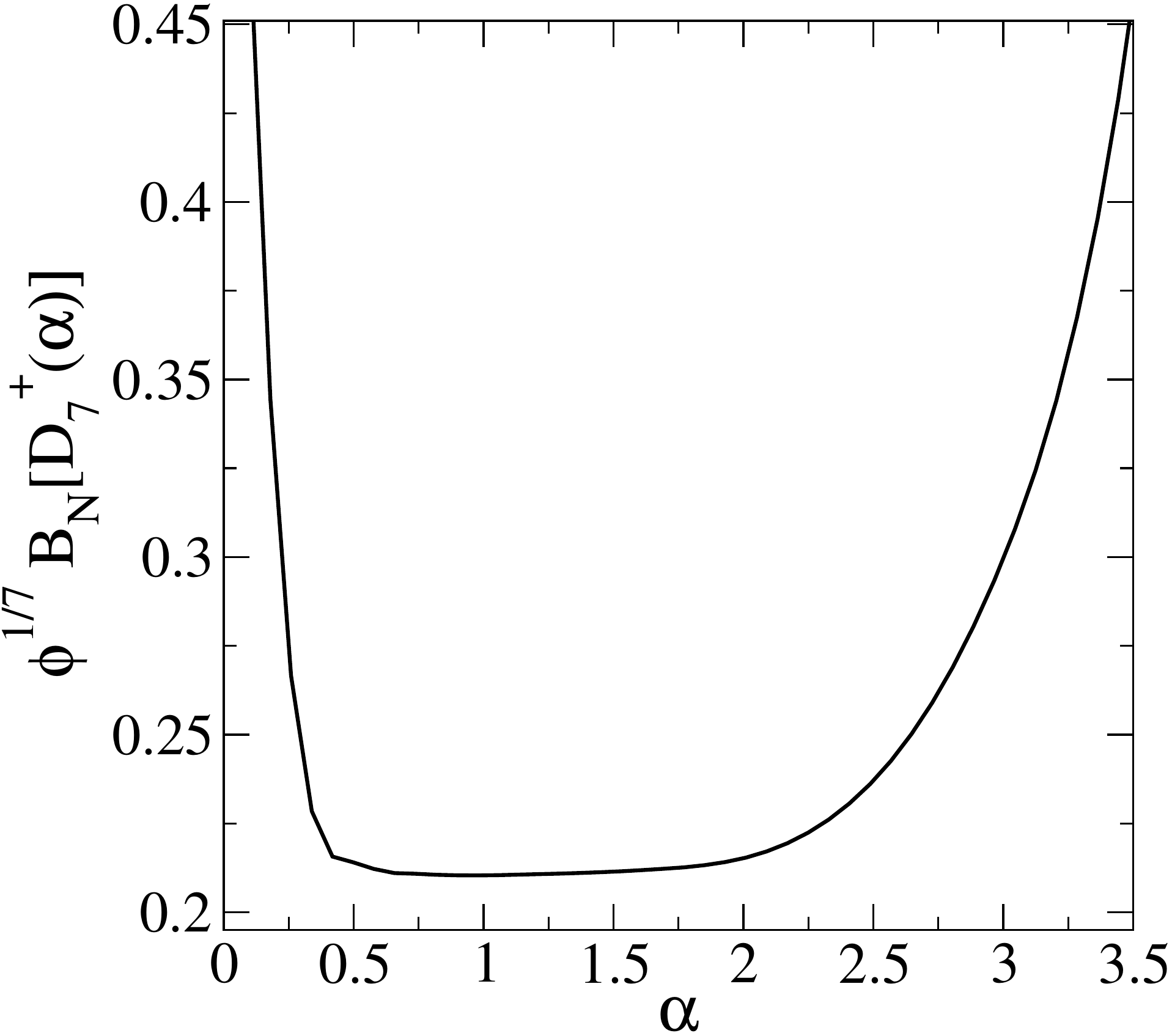}
\caption{\emph{Left:}  Schematic of the asymptotic number variance coefficients for the $D_5^+(\alpha)$ family of nonlattice packings.  The self-dual packing ($\alpha = 1$) is seen empirically to be a local maximum; $\Lambda_5^{2*}$ ($\alpha = 1/2$) is the apparent global minimum with $\Lambda_5^2$ $(\alpha = 2)$ being a local minimum.  \emph{Right:}  Corresponding schematic of the coefficients for the $D_7^+(\alpha)$ family of nonlattices; the global minimum is empirically found to be the self-dual packing.}\label{D57alpha}
\end{figure}

For $d = 5$, we have numerically found that the asymptotic number variance coefficient for $\Lambda_5^2$ is a local minimum  within the $D_5^+(\alpha)$ family; the formal dual $\Lambda_5^{2*}$ may be the global minimum (see Figure \ref{D57alpha}).  In fact, it appears that $\Lambda_5^{2*}$ may possess a lower number variance coefficient than $D_5^*$, albeit by an unusually narrow margin (see Tables \ref{latt-table} and \ref{nonlatt_table}).  Using approximately the first $15000$ terms of the theta series for both structures, we have found that $\phi^{1/5}B_N(\Lambda_5^{2*}) = 0.1906920 \pm 1.520611 \times 10^{-7}$ and $\phi^{1/5} B_N(D_5^*) = 0.1907016 \pm 1.414410 \times 10^{-7}$, where the listed errors represent standard errors from a linear regression analysis.  Since the theta series for $\Lambda_5^{2*}$ is distinct from that of $D_5^*$ and since a lattice determines a theta series (but not conversely), we expect that this difference, though small, is significant and not an artifact of numerical imprecision.  This conclusion is corroborated by the observation that the numerical values for the asymptotic number variance coefficients above are empirically constant when incorporating additional terms from the theta series.    We remark that the formally self-dual packing $D_5^+(1)$ appears to be a local maximum in the $D_5^+(\alpha)$ family with respect to the number variance coefficient.    In contrast, neither $\Lambda_7^3$ nor its formal dual appear to be extrema among the packings in the $D_7^+(\alpha)$ family; the global minimum is apparently the formally self-dual nonlattice $D_7^+(1)$ as shown in Figure \ref{D57alpha}.  However, $\Lambda_7^3$, $\Lambda_7^{3*}$, $D_7^+(1)$, and any nonlattices in between possess lower number variance coefficients than $E_7^*$ (Tables \ref{latt-table} and \ref{nonlatt_table}), suggesting that the dual of the densest Bravais lattice in a given dimension may no longer be even a local minimum with respect to the number variance in higher dimensions.  Note also that if $\Lambda_5^{2*}$ and $D_7^+(1)$ are indeed the global minima for the number variance in five and seven dimensions, respectively, then it follows that the duals of these systems should minimize the Epstein zeta function of argument $s = (d+1)/2$.  This observation is a direct result of the equivalence between the number variance and Epstein zeta function established by Torquato and Stillinger \cite{ToSt03}.  

\section{Concluding remarks}

We have extended the notion of hyperuniformity to incorporate both stochastic point processes and two-phase random heterogeneous media.  In the case of microstructures of impenetrable
spheres, an upper bound can be established on the asymptotic fluctuations in the local volume fraction in terms of the corresponding results for the local number variance
of the underlying point process.  The information contained in this estimate is particularly applicable for high-dimensional systems, where decorrelation reduces long-range 
spatial fluctuations in the system and decreases the leading-order term in the asymptotic local volume fraction variance.  Furthermore, we have provided an extensive characterization 
of hyperuniform point processes in several dimensions; our results indicate that the asymptotic coefficient $B_N$ may provide a quantitative order metric describing the extent of 
spatial uniformity.  This order metric faithfully reproduces our physically intuitive ordering of low-dimensional structures and has surprising implications for understanding order in high-dimensional systems. 
By studying several lattice families up to $d = 8$, we have shown that a Bravais lattice generally possesses a higher asymptotic number variance coefficient than
its dual.  Furthermore, in dimensions five and seven we have shown for the first time that it is possible for a nonlattice to possess a lower asymptotic number variance 
coefficient than any Bravais lattice, and the corresponding duals of these nonlattices should therefore also minimize the Epstein zeta function for argument values $s = (d+1)/2$.  

One promising direction for future research is to construct hyperuniform point patterns with targeted values of the asymptotic number variance coefficient $\phi^{1/d} B_N$ \cite{ToSt03}.  This problem 
can be expressed in terms of an optimization scheme, whereby one either seeks:
\begin{eqnarray}\label{opti}
\mbox{min}_{\forall r_{ij} \leq 2L} ~\frac{1}{L} \int_0^L \phi^{1/d} B_N(R) dR \qquad (L \rightarrow +\infty),
\end{eqnarray}
where $B_N(R)$ is given by (\ref{Bnsmall}), or attempts to minimize the square fluctuation of this objective function from a target value.  The spatial averaging in (\ref{opti}) is necessary
to smooth the small-scale fluctuations that arise from the volume-average interpretation of the number variance fluctuations, which is applicable for a single point pattern and therefore 
appropriate for the optimization scheme.  Since we have established a quantitative order metric in this paper based on the asymptotic coefficient for the number variance, it is 
feasible that one could develop new types of quasiperiodic or disordered structures with similar applications to the materials studied by Batten, Stillinger, and Torquato \cite{BaStTo08}.  

Additionally, it is an open problem to systematically construct a heterogeneous medium that is hyperuniform with respect to the local volume fraction.  
One feasible approach in this regard is to utilize so-called ``inverse construction'' algorithms \cite{YeTo98a, YeTo98b, JiStTo07, JiStTo08}.  However, this issue is complicated by the fact that not 
every heterogeneous medium can be derived from an underlying stochastic point process; in other words, it is not possible to construct an arbitrary microstructure by decorating 
some point process in the space.  One therefore seeks autocovariance functions $\chi(\mathbf{r})$ that are \emph{realizable} as heterogeneous media, meaning that there exists a binary stochastic
process $\left\{I^{(i)}(\mathbf{x}) : \mathbf{x} \in \mathbb{R}^d\right\}$ with autocovariance function $\chi$.  Several necessary but not sufficient conditions for realizability have been
given in the literature \cite{To99, To06}, and it known that numerical constructions of heterogeneous systems are sensitive to the realizability of the $n$-point probabilities \cite{JiStTo07, JiStTo08}.  Further research 
in this area is certainly warranted.  

\ack
The authors are grateful to Henry Cohn for useful discussions and for sharing the theta series data for the five- and seven-dimensional nonlattice packings.  We also thank Robert Batten
for providing the details on and configurations of classical disordered ground states.  This work was supported by the Office of Basic Energy Sciences, U.S. Department of Energy, Grant No. DE-FG02-04-ER46108.

\end{document}